\documentclass{article}

\usepackage{arxiv}
\usepackage{array}
\usepackage[utf8]{inputenc} % allow utf-8 input
\usepackage[T1]{fontenc}    % use 8-bit T1 fonts
\usepackage{hyperref}       % hyperlinks
\usepackage{url}            % simple URL typesetting
\usepackage{booktabs}       % professional-quality tables
\usepackage{amsfonts}       % blackboard math symbols
\usepackage{nicefrac}       % compact symbols for 1/2, etc.
\usepackage{microtype}      % microtypography
\usepackage{lipsum}		% Can be removed after putting your text content
\usepackage{graphicx}
\graphicspath{ {./fig/} }
\usepackage[square,numbers]{natbib}
\usepackage{multirow}
\usepackage{doi}
\usepackage{nomencl}
\usepackage[ruled]{algorithm2e}
\usepackage{float}
\usepackage{enumitem}
\usepackage{pifont}

\usepackage[most]{tcolorbox}
\usepackage{xcolor}
\usepackage{lmodern}

\usepackage{listings}
\usepackage{xcolor}
\usepackage[most]{tcolorbox}

\usepackage{caption}

\usepackage{tabularx}

\tcbset{
  colback=gray!10,     % 背景色
  colframe=gray!30,    % 边框色
  boxrule=1pt,         % 边框粗细为0（可改为1pt）
  arc=1mm,             % 圆角
  left=2mm, right=2mm, top=1mm, bottom=1mm,
  fontupper=\ttfamily, % 打字机字体
  enhanced
}

\title{AutoTRIZ: Automating Engineering Innovation with TRIZ and Large Language Models}
\date{A Preprint Version: Mar 24, 2025\thanks{An earlier version was published at DETC2024: \texttt{https://doi.org/10.1115/DETC2024-143166}}}

\author{ 
{
\hspace{1mm}Shuo Jiang} \\
	Department of Systems Engineering \\
        City University of Hong Kong, Hong Kong \\
	\texttt{shuo.jiang@cityu.edu.hk} \\
	\And
        {
        {\hspace{1mm}Weifeng Li}} \\
	School of Vehicle and Mobility \\
        Tsinghua University, Beijing, China \\
	\texttt{lwf19@mails.tsinghua.edu.cn} \\
        \And
        {
        {\hspace{1mm}Yuping Qian}} \\
	School of Vehicle and Mobility \\
        Tsinghua University, Beijing, China \\
	\texttt{qianyuping@tsinghua.edu.cn} \\
        \And
        {
        {\hspace{1mm}Yangjun Zhang}} \\
	School of Vehicle and Mobility \\
        Tsinghua University, Beijing, China \\
	\texttt{yjzhang@tsinghua.edu.cn} \\
        \And
        {
	{\hspace{1mm}Jianxi Luo}} \\
	Department of Systems Engineering \\
        City University of Hong Kong, Hong Kong \\
	\texttt{jianxi.luo@cityu.edu.hk} \\
}

\renewcommand{\headeright}{AutoTRIZ: Automating Engineering Innovation with TRIZ and LLMs}
\renewcommand{\undertitle}{}
\renewcommand{\shorttitle}{}

\begin{document}
\maketitle

\begin{abstract}
	Various ideation methods, such as morphological analysis and design-by-analogy, have been developed to aid creative problem-solving and innovation. Among them, the Theory of Inventive Problem Solving (TRIZ) stands out as one of the best-known methods. However, the complexity of TRIZ and its reliance on users' knowledge, experience, and reasoning capabilities limit its practicality. To address this, we introduce AutoTRIZ, an artificial ideation system that integrates Large Language Models (LLMs) to automate and enhance the TRIZ methodology. By leveraging LLMs’ vast pre-trained knowledge and advanced reasoning capabilities, AutoTRIZ offers a novel, generative, and interpretable approach to engineering innovation. AutoTRIZ takes a problem statement from the user as its initial input, automatically conduct the TRIZ reasoning process and generates a structured solution report. We demonstrate and evaluate the effectiveness of AutoTRIZ through comparative experiments with textbook cases and a real-world application in the design of a Battery Thermal Management System (BTMS). Moreover, the proposed LLM-based framework holds the potential for extension to automate other knowledge-based ideation methods, such as SCAMPER, Design Heuristics, and Design-by-Analogy, paving the way for a new era of AI-driven innovation tools.
\end{abstract}

% keywords can be removed

\section{Introduction}
\label{sec1}

Intuitive or structured ideation methods such as brainstorming, morphological analysis, and mind-mapping \cite{Zwicky1967, White2012, Camburn2020} have been used to aid creative ideation of human designers for concept generation. Among these, the Theory of Inventive Problem Solving (TRIZ) \cite{altshuller1999innovation} stands out as one of the most well-known approaches, widely applied for systematic innovation. TRIZ is a knowledge-based ideation methodology, which systematically structures engineering parameters, contradictions and derived from analyzing a large number of patents. This structured knowledge forms a basis by categorizing problems and solutions in a formal framework, enabling systematic problem-solving. However, the complexity of TRIZ presents significant cognitive challenges to effectively learning and applying it. In addition, the problem-solving process in TRIZ is highly dependent on the reasoning capabilities of human users. While some researchers have employed natural language processing and machine learning techniques to support specific steps within TRIZ \cite{Cascini2007,Guarino2022,Hall2022,Yan2015,Chang2023,Chou2014,Wang2023Solar,li2012framework,Berdyugina2023}, the effectiveness still depends heavily on the users’ own knowledge and proficiency with TRIZ.

Large Language Models (LLMs) such as GPT \cite{openai2023} and DeepSeek \cite{Guo2025} have not only acquired broad knowledge but also developed emergent abilities such as in-context learning \cite{Wei2022}, instruction following \cite{Wei2022} and step-by-step reasoning \cite{Wei2022}. These capabilities have been applied across various domains, including medicine \cite{Singhal2023}, chemistry \cite{Boiko2023}, mathematics \cite{Romera2024}, manufacturing \cite{Zhou2024}, robotics \cite{Mei2024}, and engineering management \cite{Jiang2025}. Recently, researchers have evaluated the capabilities of LLMs in engineering-related tasks \cite{Picard2023,Makatura2023}. These evaluations report the extensive engineering knowledge pretrained within these models and their broad applicability in engineering design and manufacturing. In the field of engineering problem-solving and idea generation, some prior research has already conducted preliminary exploration on these LLM-based methodologies \cite{Wang2023task,Han2023,Zhu2023bio,Zhu2023gen,Chen2024AskNatureNet}. However, the lack of transparency and limited control over reasoning steps during ideation often leads to divergent and insufficiently specific results, requiring multiple heuristic attempts by users to achieve desired outcomes. This process places significant demands on their domain-specific expertise. Besides, the interpretability of generated concepts remains challenging, as users obtain only the final results without understanding the ideation reasoning process.

In this work, we aim to leverage advanced reasoning capabilities and the broad knowledge of LLMs to automate TRIZ, showcasing the potential of LLMs in design automation and interpretable innovation. We have developed an LLM-based intelligent tool, AutoTRIZ, which facilitates artificial ideation for problem-solving with TRIZ-based interpretability. AutoTRIZ begins with a problem statement from the user and automatically generates a report that includes multiple solutions, strictly following the TRIZ thinking flow. To demonstrate the effectiveness of AutoTRIZ, we conducted comparative experiments with textbook cases and a detailed case study on battery thermal management system design.

Therefore, this research contributes to the growing studies about data-driven engineering design \cite{Wang2022DataDriven,Vlah2022}, generative design and innovation \cite{An2024,Jiang2023BioInspired}, and LLM-driven applications for product development \cite{Zhou2024,Zhu2023bio,Zhu2023gen,Chen2024AskNatureNet}. This paper is organized as follows: Section 2 reviews the TRIZ method, its variations, and existing research on LLMs for design and innovation. Section 3 presents the technical details of AutoTRIZ, followed by the experimental evaluation in Section 4. Section 5 provides a detailed case study to demonstrate the use procedure and effectiveness of AutoTRIZ. Section 6 offers a summary discussion, and Section 7 discusses the limitations and future work. Finally, Section 8 concludes the paper.

\section{Related Work}
\label{sec:sec2}

\subsection{TRIZ}

TRIZ is a knowledge-based systematic approach of inventive problem solving, developed in the 1960s by Genrich S. Altshuller and his colleagues \cite{altshuller1999innovation}. Through a thorough analysis of over 40,000 patents, Altshuller and his collaborators identified repeated patterns of innovation and underlying innovative principles within these documents. By inductively analyzing these patterns, they proposed a comprehensive problem-solving framework, applying selected inventive principles for ideation. Since then, the TRIZ has been developed continually and some modern TRIZ databases rely on analysis of over 2 million patents. It has been widely applied in industries, research, and education with notable influence in many fields, such as energy, electrical, automotive industries, and mechanical engineering \cite{Spreafico2016}.

The TRIZ toolkit contains a series of theories and tools that cover all aspects of problem understanding and solving, including trimming method, evolution trends, and 76 standard solutions \cite{altshuller1999innovation}. In this paper, we focus on the inventive principles, which represent the basic reasoning logic behind TRIZ. Figure 1 shows the overview of its framework, which contains four steps: 

\begin{enumerate}[label=(\arabic*)]
  \item \textbf{Identify the specific problem.} Clearly define the engineering problem by extracting key details from the given context. This includes analyzing the functional requirements, constraints, and existing inefficiencies within the system.
  \item \textbf{Transform the specific problem into a general problem.} This transformation is achieved by mapping the problem onto TRIZ’s engineering parameters. By doing so, contradictions within the system can be identified, involving an improving feature and a worsening feature. These contradictions form the basis for TRIZ problem-solving.
  \item \textbf{Use the contradiction matrix to find relevant inventive principles.} The contradiction matrix consists of 39 improving features and 39 worsening features, forming a 39 × 39 grid. Each cell in this matrix suggests the most commonly used inventive principles (drawn from TRIZ’s 40 inventive principles) that may help resolve the identified contradiction.
  \item \textbf{Apply the selected inventive principles to generate solutions.} The inventive principles serve as guidelines for problem-solving, helping to generate specific solutions that address the contradiction. Depending on the nature of the problem, analogical thinking may be used to draw insights from known solutions in other fields, or general solutions may be adapted to solve the specific problem at hand.
\end{enumerate}

\begin{figure}[H]
	\centering
	\includegraphics[width=11cm]{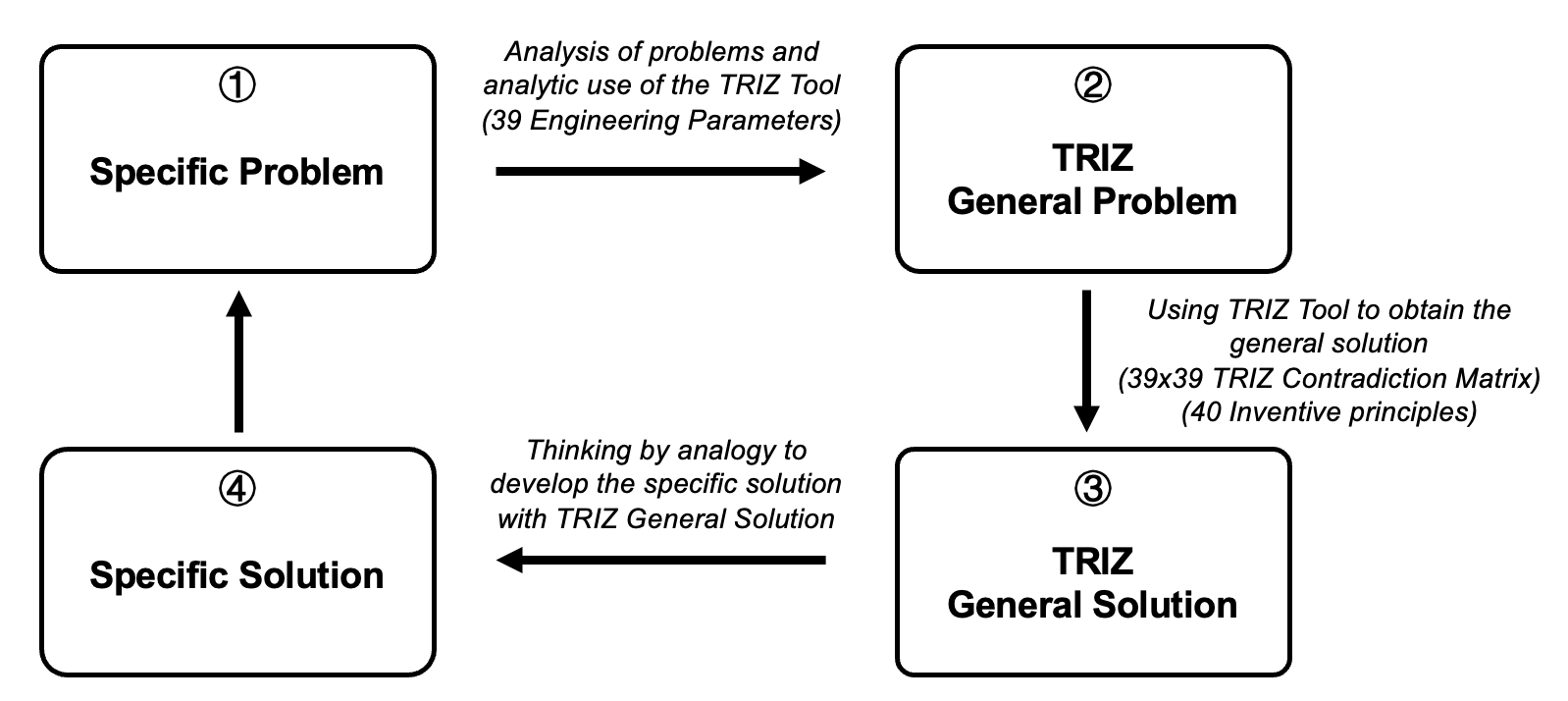}
	\caption{Four steps for problem-solving using TRIZ}
	\label{fig:fig1}
\end{figure}

Although TRIZ has demonstrated its effectiveness, it still suffers drawbacks that hinder its practical applications. For instance, the complexity of TRIZ resources and concepts poses cognitive challenges to effectively learning and applying it, particularly for non-experts. Additionally, the efficacy of TRIZ is heavily constrained by the users’ reasoning capabilities and prior knowledge acquired.

Recent advancements in machine learning, natural language processing and computational intelligence have been applied in conjunction with TRIZ \cite{Cascini2007,Guarino2022,Hall2022,Yan2015,Chang2023,Chou2014,Wang2023Solar,li2012framework,Berdyugina2023}. These efforts aim to automate the TRIZ reasoning process, reducing its difficulty. For instance, Cascini and Russo \cite{Cascini2007} developed the PAT-ANALYZER system that can analyze patent texts and automatically extract the contradictory information underlying the innovation for TRIZ. Similarly, Guarino et al. \cite{Guarino2022} proposed the PaTRIZ, combining the Bidirectional Encoder Representations from Transformers (BERT) and Conditional Random Fields (CRF) for word-level patent analysis and TRIZ contradiction mining. Hall et al. \cite{Hall2022} proposed an approach that uses topic modeling and unsupervised machine learning to map TRIZ inventive principles to individual patents and detect the novelty. Yan et al. \cite{Hall2022} proposed IngeniousTRIZ, an automatic ontology-based system for solving inventive problems. It first requires users to choose an initial TRIZ knowledge source as an abstract solution, then uses semantic similarity computation to select other relevant TRIZ knowledge and generates final solutions through ontology inference. Chang et al. \cite{Chang2023} proposed a user-centric smart product-service system that integrates TRIZ, service logic, and product architecture theories. This system employs a user journey map-based method to analyze user needs and map them into TRIZ engineering parameters through semantic relations, ultimately generating conceptual solutions using the identified relevant TRIZ inventive principles. Chou \cite{Chou2014} introduced an ideation framework for developing new product concepts using TRIZ, concept mapping and fuzzy linguistic evaluation. Wang et al. \cite{Wang2023Solar} leveraged Random Forest (RF) algorithm to mine the relationship among customer needs, TRIZ engineering principles and inventive principles to support engineering innovation. Li et al. \cite{li2012framework} proposed an approach that leverages natural language processing techniques to assess patent innovations according to the level of invention as defined in TRIZ. Berdyugina and Cavallucci \cite{Berdyugina2023} proposed a text mining-based methodology for automatically extracting inventive information from texts to formulate an inventive problem into TRIZ engineering parameters. However, most of these works utilize algorithms to improve specific steps of the TRIZ process. They still require innovators to dedicate much time and effort to extensive reasoning. Employing these methods does not directly assist users throughout the entire process, from analyzing a problem to creating practical solutions. In this paper, we aim to harness LLMs to automate the whole TRIZ reasoning process and minimize the cognitive requirements for users during its application.

\subsection{Large Language Models for Engineering Design and Innovation}

Over the past years, many data-driven approaches have utilized machine learning and deep learning techniques to augment design and innovation \cite{Luo2022, Jiang2022}. Evolved from deep learning and pre-trained language models, LLMs typically refer to Transformer-based models that contain hundreds of billions of parameters for processing and generating natural language texts \cite{Zhao2023}. They are trained on extremely large-scale corpora, enabling them to acquire a wide range of knowledge and capabilities, including understanding context, generating coherent text, and step-by-step reasoning \cite{Wei2022}. Some research has already explored the application of LLMs in engineering design and innovation within specific fields, including microfluidic devices \cite{Nelson2023}, robotics \cite{Stella2023}, and the user interface of webpages \cite{Li2023}. However, most of these early efforts primarily utilize conversational interactions, such as those facilitated by ChatGPT Interface \cite{Nelson2023}, to engage in the innovation process. Meanwhile, with the development of LLMs, there has been an increase in efforts to create LLM-driven methods and tools to offer more generalized innovation assistance and directly support users in rapid ideation.

For instance, several studies have harnessed LLMs for processing vast amounts of design documentation, representing designs in specific forms, and identifying user needs for product development \cite{Wang2023task,Han2023,Qiu2023,Chen2024Controllable}. Han et al. \cite{Han2023} introduced an LLM-based attribute-sentiment-guided summarization model to extract user needs from online product reviews. Qiu et al. \cite{Qiu2023} applied a transformer-based language model to distill design-related knowledge from extensive reports and documents. Moreover, Chen et al. \cite{Wang2023task,Chen2024Controllable} utilized LLMs to decompose conceptual design tasks into Function-Behavior-Structure (FBS) formats, assisting users in ideation across different aspects.

Recent studies have developed tools and methodologies utilizing LLMs to aid the design process, enhance human-computer collaborative innovation, or directly produce innovative concepts for users \cite{Zhu2023bio,Zhu2023gen,Chen2024AskNatureNet}. Siddharth and Luo \cite{Siddharth2024} proposed a method to extract explicit engineering design facts from patent artifact descriptions to support retrieval-augmented generation of LLMs in the design process. Ding et al. \cite{Ding2023} systematically explored LLMs’ potential to boost cross-domain analogical creativity. Chen et al. \cite{Chen2024AskNatureNet} proposed AskNatureNet, a divergent design ideation tool based on biological knowledge. They first constructed a specialized knowledge graph using LLMs and biological texts from the AskNature database, and then developed retrieval and mapping methods to support analogical reasoning and its applications. Huang et al. \cite{Huang2023} proposed CausalMapper, a system that combines LLMs with causal mapping to explain the connections between problems and solutions. Ma et al. \cite{Ma2023,Ma2025} evaluated the differences between LLM-generated and crowdsourced design solutions through multiple perspectives, including human expert evaluations and computational metrics. Zhu and Luo \cite{Zhu2023gen} presented GPT-based models with domain-specific tuning and task-specific learning to generate original and practical design concepts. They also applied their approach to automating bio-inspired design concept generation \cite{Zhu2023bio}. Chen et al. proposed TRIZ-GPT \cite{Chen2024TRIZGPT}, which explores and evaluates different reasoning strategies of LLMs for the TRIZ-based problem-solving process. Lee et al. \cite{Lee2024} integrated LLMs, retrieval-augmented generation (RAG), and TRIZ methodology to automate the ideation process in eco-design. Their proposed Eco-Innovate Assistant provides users with eco-innovative solutions, complemented by design sketches.

While these recent idea-generation approaches effectively leverage the reasoning capabilities of LLMs, they still require varying degrees of guidance and human intervention to ensure that solutions align with specific design needs. Managing the problem-solving process to ensure that solutions are both innovative and practical, as well as understanding the reasoning behind generated innovative solutions, remains a challenge. In this study, we address this issue by integrating TRIZ with LLMs, presenting AutoTRIZ as an end-to-end tool that follows the TRIZ reasoning steps to generate inventive solutions.

\section{AutoTRIZ}
\label{sec3}

In this section, we introduce AutoTRIZ, an artificial ideation tool that automates TRIZ with LLMs. The architecture of AutoTRIZ is depicted in Figure 2. Overall, AutoTRIZ takes a problem statement from the user as its initial input and automatically generates a solution report. The report includes detailed information about the TRIZ reasoning process and generated solutions guided by identified TRIZ inventive principles. While AutoTRIZ automates the entire reasoning process, it still follows the core TRIZ logic of “from specific to general” and “general to specific,” leveraging LLMs to facilitate these steps efficiently. Within AutoTRIZ, we have defined a four-step reasoning flow based on the classic TRIZ workflow. The system includes an inner fixed knowledge base consisting of three segments related to TRIZ details, enabling controlled reasoning. It is noteworthy that our focus is on controlling the entire problem-solving reasoning process while remaining open to the knowledge used in ideation. The problem-related knowledge applied during the problem-solving process comes from the LLM pre-training stage on the large-scale corpus.

\begin{figure}[H]
	\centering
	\includegraphics[width=16cm]{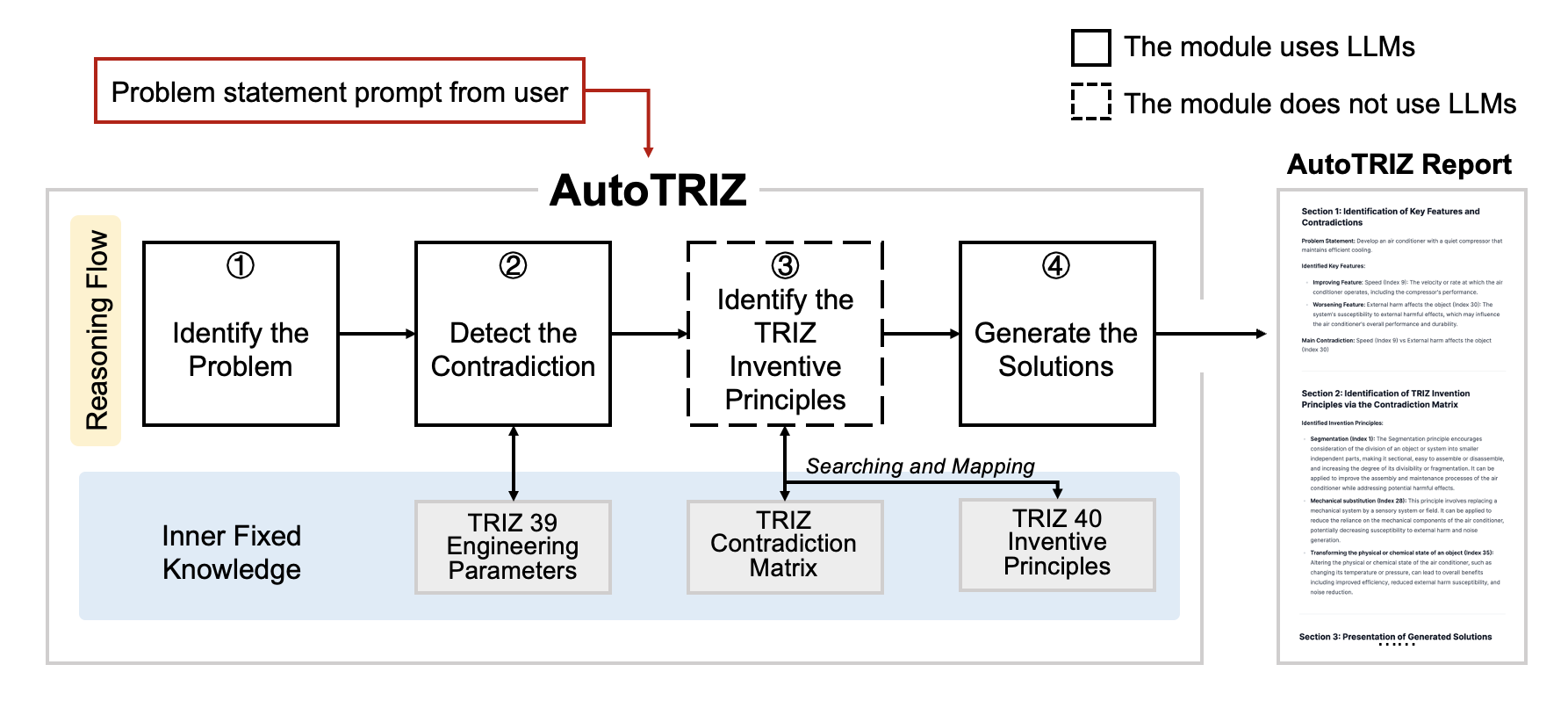}
	\caption{The framework of AutoTRIZ}
	\label{fig:fig2}
\end{figure}

\subsection{Controlling the TRIZ Reasoning Flow}

To ensure that the system strictly follows the TRIZ reasoning process, we have configured AutoTRIZ with four modules, each corresponding to the four steps in TRIZ. As depicted in Figure 2, Modules 1, 2, and 4, outlined by solid-line frames, are driven by LLMs, whereas Module 3, outlined by a dashed-line frame, is controlled by predefined functions without using LLMs. Specifically, we exploit the instruction-following capabilities of LLMs for backend reasoning control. In each module incorporating LLMs, relevant instructions are engineered into the input as system and assistant prompts.

Specifically, in Module 1, AutoTRIZ identifies the problem to be solved from user input $U$ and converts it into descriptive text $T$. Ideally, we hope that the content entered by the user is a clear problem statement. However, user inputs may include additional information such as scenario descriptions, background details, and some redundant information. Therefore, in this module, AutoTRIZ is designed to identify and extract information related to the problem and then reorganize it into clear and concise text, expressed as:
$$T=f_{1}{(U)}$$

In Module 2, AutoTRIZ receives the processed problem description $T$ and detects its engineering contradiction, which is represented by a space constructed from two out of the 39 engineering parameters $\{P_{1},P_{2},...,P_{39}\}$. Let $C$ represent the identified contradiction. AutoTRIZ learns all the engineering parameters from the inner knowledge base at this stage. The outputs of this module are presented in a structured format (i.e., the indexes of the improving and worsening features). This process can be represented as:
$$C=\{P_{improve},P_{worsen}\}=f_{2}{(T)}$$

It is important to note that for the same problem statement $T$, the identified contradiction $C$ may differ with each execution of this module. On the one hand, a single problem may encompass multiple contradictory pairs, yet our system is designed to identify only one contradiction. On the other hand, there is an inherent randomness in the content generation by LLMs. In Section 4, we will conduct experimental investigations to examine the effectiveness of contradiction identification and the consistency of the outputs.

Once the contradiction is identified, Module 3 searches the contradiction matrix $CM$ to find the indexes of corresponding inventive principles and returns their descriptions. Formally, the set of inventive principles $I$ corresponding to $C$, is found by querying the contradiction matrix, expressed as:
$$I=\{I_{1},I_{2},...,I_{40}\}=f_{3}{(CM,C)}$$

Following this, Module 4 synthesizes the original problem description $T$, the identified engineering contradiction $C$, and the inventive principles $I$, to generate the set of final solutions $S$, expressed as:
$$S=f_{4}{(T,C,I)}$$

LLMs can generate complex structured data, such as in HTML and \LaTeX formats \cite{Tang2023}. In AutoTRIZ, we harness this capability to integrate all generated content and directly produce a reader-friendly problem-solving report in a structured format. We have engineered the format template directly into module 4, enabling it to output documents formatted in \LaTeX. In practice, the template for the report generation can be adjusted as needed to suit specific requirements.

\subsection{Learning from the Fixed Knowledge Base}

AutoTRIZ acquires the necessary information to learn the prior knowledge of TRIZ, enabling it to handle various engineering problems. We have employed a static knowledge base that interacts with the modules described above, empowering AutoTRIZ to master and apply the relevant TRIZ knowledge.

In AutoTRIZ, the internal fixed knowledge base includes three main components: (1) the TRIZ 39 Engineering Parameters \cite{altshuller1999innovation}, (2) the TRIZ Contradiction Matrix \cite{altshuller1999innovation}, and (3) the TRIZ 40 Inventive Principles \cite{altshuller1999innovation}. Notably, the contradiction matrix here is identical to the traditional TRIZ contradiction matrix. The knowledge about engineering parameters and inventive principles includes titles and detailed descriptions for each entry. 

For example, the first engineering parameter \texttt{\{Weight of moving object (CP1)\}} is described in the following format:
\begin{tcolorbox}[breakable]
\texttt{[INDEX] 1 [TITLE] Weight of moving object} \\ 
\texttt{[DESCRIPTION] The mass of the object in a gravitational field, essentially the force that the body exerts on its support or suspension.}
\end{tcolorbox}

Similarly, the first inventive principle \texttt{\{Segmentation (IP1)\}} is described as follows:
\begin{tcolorbox}[breakable]
\texttt{[INDEX] 1 [TITLE] Segmentation} \\ 
\texttt{[DESCRIPTION] The Segmentation principle encourages consideration of the division of an object or system into smaller independent parts, making it sectional, making it easy to assemble or disassemble, and increasing the degree of its divisibility or fragmentation.}
\end{tcolorbox}

The 39 engineering parameters are configured into Module 2 as assistant information. The backend LLMs learn instructions and the output parameter space through in-context learning, enabling zero-shot reasoning. Regarding inventive principles, only selected items are exposed to the system based on their position in the contradiction matrix. This process resembles LLMs’ Retrieval Augmented Generation (RAG) technology \cite{Siddharth2024,Lewis2020}. By retrieving additional information related to the query from external databases, RAG incorporates these external texts into LLM prompts to address the hallucination problem, leading to better generation \cite{Lewis2020}. In our system, the problem-solving process involves precise search-augmented generation, effectively bridging the gap between the prior TRIZ knowledge from experts and the generic reasoning capabilities of LLMs derived from large-scale pre-training.

\subsection{System Implementation}

We have developed a web-based tool for public users to test and use AutoTRIZ, available at: \href{https://www.autotriz.ai}{https://www.autotriz.ai/}. Figure 3 shows the user interface of the tool. Throughout the deployment of this tool and all experiments and case studies presented in this study, we utilized \textit{GPT-4} (Version: 20231106, the state-of-the-art model at the time this work was done) as the backend LLM. However, it is essential to note that since the proposed AutoTRIZ is a general framework, the backend LLM can be replaced with any other closed-source LLM (e.g., Claude) or open-source LLM (e.g., DeepSeek). For the TRIZ knowledge base used in AutoTRIZ, we adopt the TRIZ definitions and descriptions in an engineering design textbook \cite{Childs2013}. To further explore the impact of different LLMs on AutoTRIZ’s reasoning and solution quality, future studies could systematically compare multiple LLMs within this framework.

\begin{figure}[H]
	\centering
	\includegraphics[width=16cm]{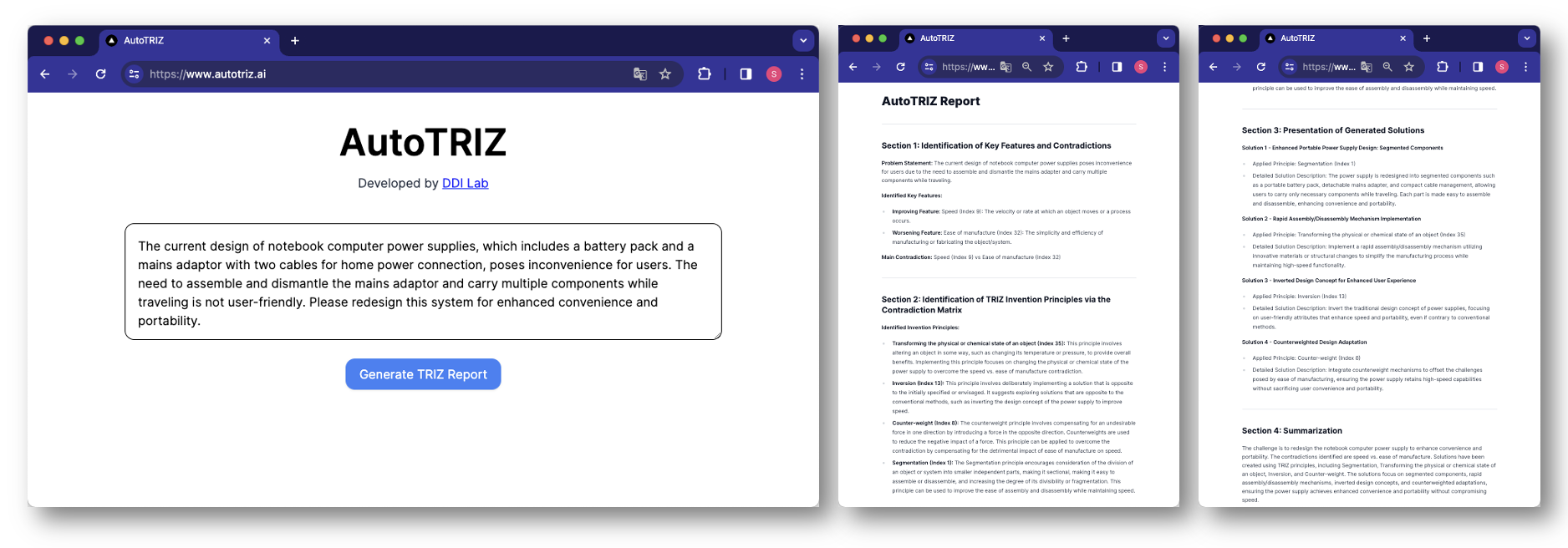}
	\caption{AutoTRIZ web-based tool}
	\label{fig:fig3}
\end{figure}

\section{Experimental Evaluation}
\label{sec:sec4}

In this section, we evaluate the effectiveness of the proposed AutoTRIZ through quantitative experiments and comparative studies with textbooks. Specifically, we collected several engineering cases analyzed by human experts from TRIZ textbooks, constructing a case base. Then, we explored the system's consistency in identifying engineering contradictions and its alignment with human analysis. Finally, we selected a specific problem from the case base and then compared and discussed the solutions generated by AutoTRIZ against the results of human experts.

\subsection{Constructing the TRIZ Case Base}

To evaluate the performance of AutoTRIZ, we first constructed a case base containing TRIZ problem-solving cases developed by human experts. Initially, we gathered several TRIZ-related textbooks, some of which are focused on general design innovation, while others are specifically about TRIZ. From 7 of these textbooks \cite{altshuller1999innovation,Childs2013,Orloff2006,Orloff2012,Savransky2000,Silverstein2008,Fey2005}, we collected 10 initial cases. The selection criteria include: (1) the content of the case contains all elements of the TRIZ reasoning process, including problem description, contradiction identification, inventive principle positioning, and solutions; (2) the problem is defined clearly and comprehensively; (3) the cases do not contain similar problems. The selection criteria include: (1) the content of the case contains all elements of the TRIZ reasoning process, including problem description, contradiction identification, inventive principle selection, and solutions; (2) the problem is defined clearly and comprehensively; (3) the cases do not contain similar problems. Two of the co-authors of this study have checked all cases to ensure accuracy. They are then stored in JSON format. For more details on collected cases, readers may refer to our GitHub repository \footnote{\href{https://github.com/shuojiangcn/AutoTRIZ-Repository/}{https://github.com/shuojiangcn/AutoTRIZ-Repository/}}
.
The initial 10 cases cover various domains, including environmental engineering, transportation, manufacturing, material science, aerospace technology, etc. Given the need for detailed TRIZ reasoning steps in each case, we prioritized depth and clarity over quantity to ensure a rigorous evaluation of AutoTRIZ’s problem-solving process. The evaluation of these cases can serve as a preliminary benchmark, enabling users to understand and get familiar with the usage protocol and performance of AutoTRIZ. In the future, we will continue to expand the case base for more robust statistical testing. Beyond serving experimental purposes in this study, the curated case base can also store the results generated by users with AutoTRIZ, making it scalable. As the size of the base expands, we can also explore the interaction between the reasoning module and the existing case base, enabling AutoTRIZ's innovative capabilities to be scalable.

\subsection{Assessing the Contradiction Identification}

Detecting contradictions is an essential step in the entire TRIZ problem-solving process. Accurate identification of the contradictions within a problem can effectively assist AutoTRIZ in recommending the appropriate inventive principles for the following steps. Within LLMs, randomness is naturally incorporated into the text generation process. These models often use sampling methods or temperature adjustments to control the generation process, leading to various possible outputs rather than repeating the same response every time. Because of this inherent variability, LLMs may suffer from instability during inference. As a result, some LLM-based agents adopt self-consistency techniques that create several reasoning paths and then perform an ensemble on all generated answers, selecting the most consistent one through majority voting \cite{Wang2023self}. However, in traditional TRIZ, analyzing the same problem from different perspectives can yield different possible contradictions. The stochastic nature of LLM-based generation can help increase the diversity of generated ideas \cite{Ma2025}. Based on this, we maintain the same setting of producing a single contradiction in each entry.

For each given problem statement, we performed the analysis 100 times, resulting in 100 pairs of identified parameters. Then, we counted all results and calculated their respective proportions. In cases of high consistency, a particular contradiction could be dominant. In some cases, one parameter in the contradiction may have higher certainty, leading to more dispersed results.

We used information entropy as the uncertainty score, where a smaller entropy value indicates greater confidence in the model's output. The information entropy metric is widely used for uncertainty measurement \cite{Zhang2019}. Given a probability distribution $X$ generated by the model, we can calculate the entropy by: $H(X) = -\sum_{i=1}^{n} P(x_i) \log_2 P(x_i)$, where $P(x_i)$ represents the probability of occurrence of the outcome, and $n$ is the number of different possible outcomes of $X$. Since the number of runs is 100, the entropy value ranges from 0 to 6.64, where a higher value indicates lower consistency.

Furthermore, we examined the overlap between AutoTRIZ’s detection and the analysis results of human experts from textbooks, categorizing them into three scenarios: complete match, half match, and no match. It is important to note that since human expert analysis also includes subjectivity and bias, it cannot be considered a golden standard. The main purpose of this experiment is to showcase and quantitatively compare AutoTRIZ against human uses of TRIZ.

Figure 4 shows the experimental results, where the bar chart of each case illustrates the top 3 detections by proportion. In the chart, green bars represent \textit{complete match}, blue bars indicate \textit{half match}, and yellow bars denote \textit{not match}. The table in the bottom right corner shows the entropy of each case and whether the top 3 detections hit the reference from textbooks, with symbols ($\checkmark$, \underline{$\checkmark$}, \ding{55}) indicating \textit{complete match}, \textit{half match}, and \textit{not match}, respectively.

Overall, 7 out of 10 cases match or half-match the textbook’s analysis within the top 3 detections, indicating that AutoTRIZ's inference overlaps with the human experts’ results to a certain degree. A minority of the cases show relatively higher consistency (cases 5, 6, 7, 8), where the proportion of the top 1 detection is significantly higher than the other detections, including two complete match detections. For these cases, utilizing self-consistency may be beneficial to enhance performance. For other cases, the experimental results show greater diversity, indicated by higher information entropy. By examining the content of the top 3 detections of contradiction for each case, we observe that for almost all cases, one parameter is fixed while the other varies. 

Moreover, when using the textbook’s analysis as a reference, a pattern emerges across all cases where outputs with higher probabilities (within the top 3 detections) show a better match in alignment. These findings can serve as the initial benchmark for assessing the performance of AutoTRIZ’s contradiction identification. As the case base expands in the future, we can explore these patterns in a more fine-grained way with greater statistical significance. For example, we can examine the differences between various themes, leveraging techniques such as self-consistency reasoning in conjunction with the identified patterns to improve overall performance.

\begin{figure}[H]
	\centering
	\includegraphics[width=14cm]{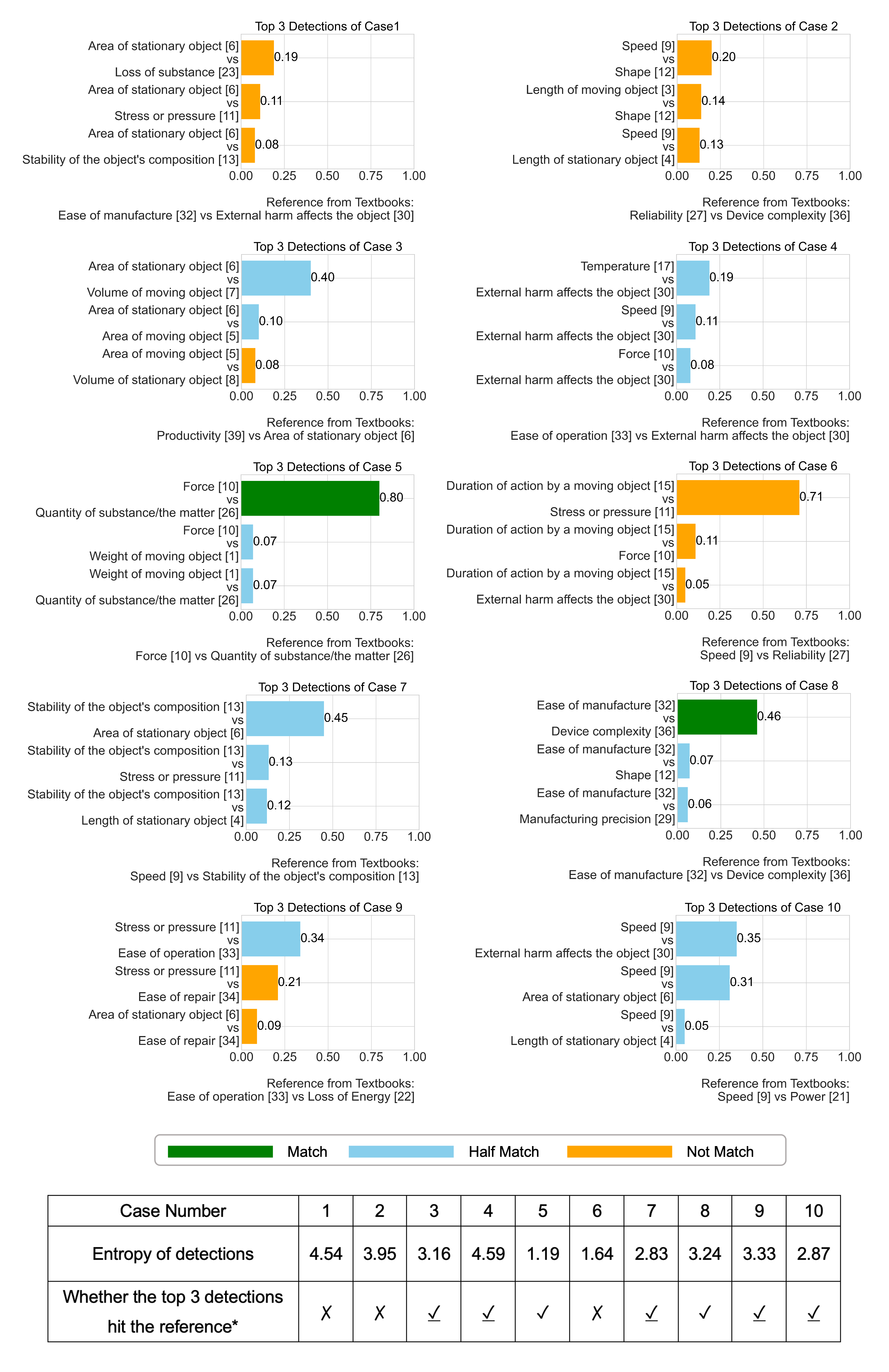}
	\caption{Experimental results about contradiction detection}
	\label{fig:fig4}
\end{figure}

\newpage
\subsection{Comparing AutoTRIZ and Human Expertise}

In this section, we select one of the collected cases (Case 7: Metal Shot Transport in Plastic Piping) to compare AutoTRIZ's generated report with experts’ analysis results from the textbook. The reasons for choosing this case are two-fold: (1) This case exhibits relatively high consistency in identifying engineering contradictions, with one dominant outcome (refer to Figure 4); (2) The top 3 detections of contradiction are all half-match with the reference. This ensures a certain degree of reliability while allowing the distinction between the subsequent reasoning paths of AutoTRIZ and humans.

Here is the original problem statement\cite{Savransky2000}:
\begin{tcolorbox}[breakable]
\texttt{We are faced with a challenge involving the pneumatic transportation of metal shots through a system of plastic piping originally intended for plastic pellets. The transition to metal shots, despite their advantages for production purposes, has led to significant wear and damage, particularly at the pipe's elbows. This issue arises from the incompatibility between the metal shots and the existing plastic elbow design. The task is to identify and implement a solution that resolves this conflict, ensuring the system's durability and effectiveness for transporting metal shots.}
\end{tcolorbox}

In the textbook, the given improving parameter is \texttt{\{Speed (CP9)\}}, and the worsening parameter is \texttt{\{Stability of the object's composition (CP13)\}}. According to the contradiction matrix, the author selects \texttt{\{Mechanical Substitution (IP28)\}} from the corresponding inventive principles. Applying this principle, the author describes the solution as \textit{placing a magnet at the elbow to bind the metal shots to a plastic material, creating a blanket of shots that absorb the energy}.

Figure 5 shows the problem-solving report generated by AutoTRIZ, containing the reasoning process and solutions. Firstly, we can see that AutoTRIZ simplifies the original problem statement, identifying the main issue that needs to be addressed. Regarding the identification of contradictions, AutoTRIZ diverges from human expertise. Both AutoTRIZ and the textbook’s analysis consistently recognize the \texttt{\{Stability of the object's composition (CP13)\}} as the worsening feature. However, concerning the improving feature, AutoTRIZ detects \texttt{\{Area of stationary object (CP6)\}}, while the textbook's analysis considers it to be \texttt{\{Speed (CP9)\}}. From the original problem statement, we can see that the critical issue is avoiding wear on the plastic elbows by the metal shots to ensure durability, which indicates that one of the engineering parameters involves stability. However, another dimension is not directly mentioned, leading to various possible interpretations. AutoTRIZ reasons that the surface area needs improvement to withstand the impact and wear of the metal shot, while the expert considers speed as the system’s top priority. These two analyses highlight different needs, thereby guiding subsequent innovative directions differently.

In the textbook's analysis, the author selected a single inventive principle \texttt{\{Mechanical Substitution (IP28)\}} and created a solution by positioning a magnet at the piping's elbow, which magnetically attaches metal shots to the plastic, forming an energy-absorbing layer. This approach represents a direct and effective innovation. However, based on the identified engineering contradiction, TRIZ could yield four inventive principles (i.e., \texttt{\{Mechanical Substitution (IP28)\}}, \texttt{\{Homogeneity (IP33)\}}, \texttt{\{Segmentation (IP1)\}}, \texttt{\{Mechanical Vibration (IP18)\}}). Some principles may be challenging to apply, as the outcomes are directly influenced by the users’ reasoning ability, experience, and familiarity with TRIZ materials. This step also requires the most human effort in TRIZ. By comparison, AutoTRIZ can effectively overcome this issue. After identifying the contradiction (\texttt{\{Area of stationary object (CP6)\}} vs \texttt{\{Stability of the object's composition (CP13)\}}), AutoTRIZ returns two inventive principles from the contradiction matrix (i.e., \texttt{\{Extraction (IP2)\}}, \texttt{\{Strong Oxidants (IP39)\}}). AutoTRIZ applies each principle and generates a novel solution. Both proposed solutions demonstrate feasibility and innovation. Solution 1 implements a physical alteration to prevent direct contact between the metal shots and the piping. Solution 2, integrating 'Strong oxidants', involves a surface treatment to improve the piping's durability against metal shots through a protective coating.

In summary, both the textbook's solution and the solutions automatically generated by AutoTRIZ are practical, originating from different inventive principles and leading to different results. In the previous section, we performed 100 trials on each case for contradiction detection. In this section, we randomly selected one trial's solutions to compare and discuss with experts’ analysis results from the textbook. We only randomly chose one solution report to demonstrate AutoTRIZ’s capability because the complete reports are too lengthy and complex to visualize and present. In future work, we will also seek computational evaluation methods and statistical metrics \cite{Regenwetter2023} regarding the quality of generated solutions.

\begin{figure}[H]
	\centering
	\includegraphics[width=10cm]{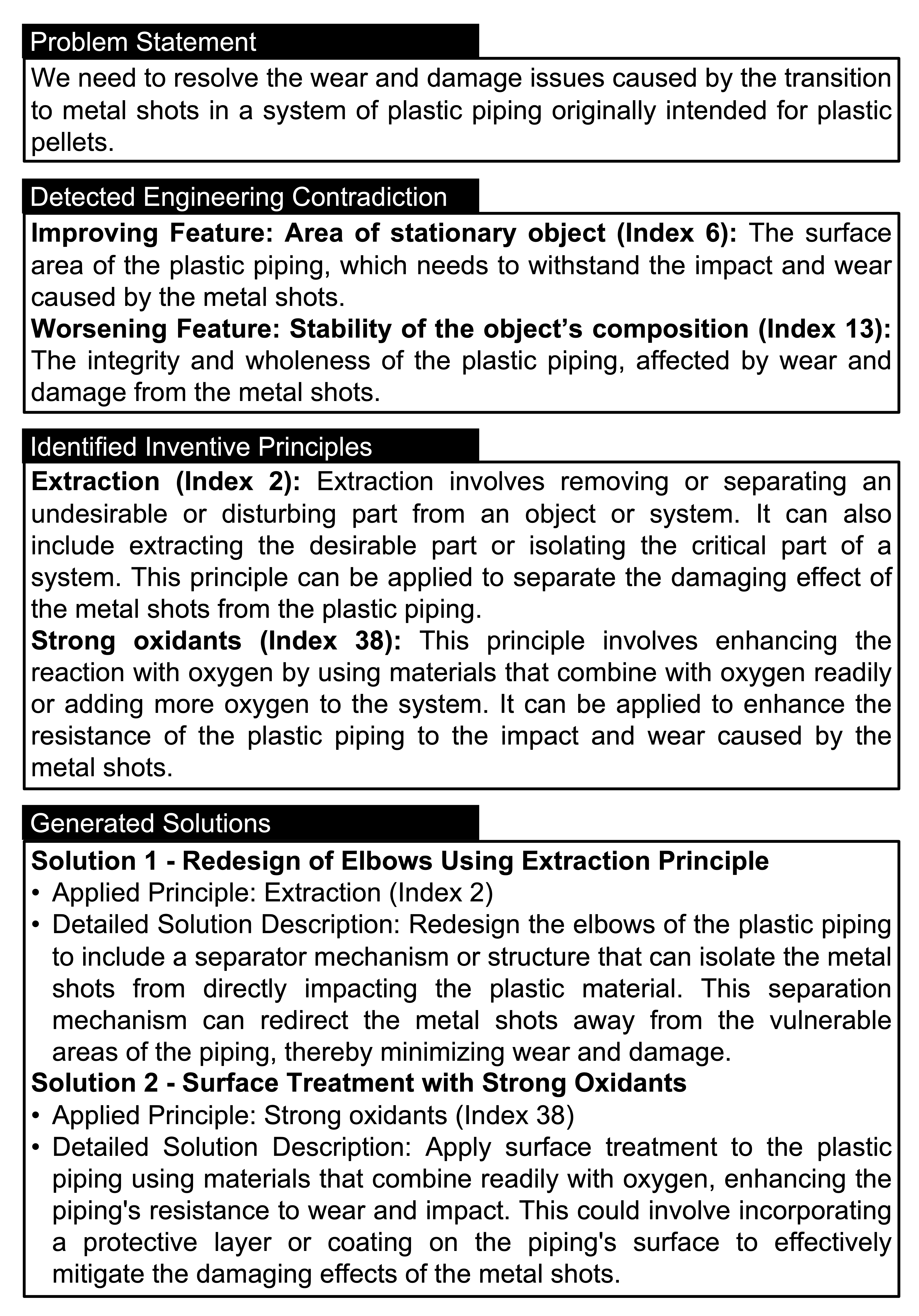}
	\caption{The AutoTRIZ-generated solution report for the problem of metal shot transport in plastic piping}
	\label{fig:fig5}
\end{figure}

\section{Case Study: Battery Thermal Management System of Electric Vehicle}
\label{sec:sec5}

\subsection{Problem Background}

With the development of Urban Air Mobility (UAM), power batteries are utilized in electric vertical take-off and landing (e-VTOL) vehicles \cite{Yang2021}. Meanwhile, the existing limitation of the driving range of Electric Vehicles (EV) has significantly increased the demand for battery fast-charging technologies \cite{Thakur2023}. The charge-discharge rate of power batteries is expected to grow continuously, leading to a significant rise in heat generation rate. This could cause thermal runaway in the batteries, potentially leading to safety incidents. Power batteries constitute 20-30\% of the total vehicle weight \cite{Belingardi2023}. As future e-VTOLs and EVs evolve with more demanding performance and range requirements, enhancing power battery systems' energy density and power density becomes increasingly crucial. Traditional air-cooling and liquid-cooling methods could not simultaneously meet the multi-objective requirements of controlling temperature rise and achieving lightweight design. These use needs and functional requirements demand innovative designs of Battery Thermal Management Systems (BTMS).

Recent research in this field suggests that heat pipe-based schemes have the potential to simultaneously meet the requirements for lightweight design and temperature control \cite{Weragoda2023}. Heat pipes utilize the principle of phase change, where the heat transfer medium evaporates and condensates at the heat absorption and dissipation sections, respectively, facilitating the absorption and release of heat. This also allows for high heat flux density with minimal temperature differences \cite{Veedu2004}. Most current studies employ copper-sintered heat pipes embedded within battery modules \cite{Smith2018,Wang2015,Liang2018,Liang2019}. However, these designs lead to insufficient contact between the heat pipes and the batteries, and the heat pipes are not fully utilized for heat transfer capabilities, which affects cooling efficiency. Additionally, the complexity of these system designs presents challenges for the subsequent operation and maintenance of the battery packs.

In this section, we utilize the proposed AutoTRIZ as a Computer-Aided Innovation (CAI) Tool to solve this problem. We focus on heat pipes and 21700 cylindrical Li-ion batteries as the foundational components of the entire BTMS, employing AutoTRIZ to directly generate design solutions for this issue and obtain innovative suggestions. The goal is to design a BTMS that meets the requirements for a lightweight structure and effective temperature control.

\subsection{BTMS Design with the Assistance of AutoTRIZ}

First, we formulate the engineering problem described above as the following statement:
\begin{tcolorbox}[breakable]
\texttt{The current lack of specifically designed heat pipes for BTMS results in complex heat transfer structures and suboptimal cooling efficiency. Please propose a novel heat pipe-based BTMS solution. The design should ensure full and direct contact between the heat pipe and the battery surface to maximize the utilization of the heat pipe's high heat transfer capability. Additionally, the solution must meet the high heat dissipation demands of the battery under high discharge rate conditions.}
\end{tcolorbox}

To statistically identify the engineering contradictions of this problem, we used AutoTRIZ to conduct 100 trials with the same input information, thus generating 100 solution reports. Figure 6 shows the frequency distribution of all identified contradictions that appeared at least three times.

\begin{figure}[H]
	\centering
	\includegraphics[width=10cm]{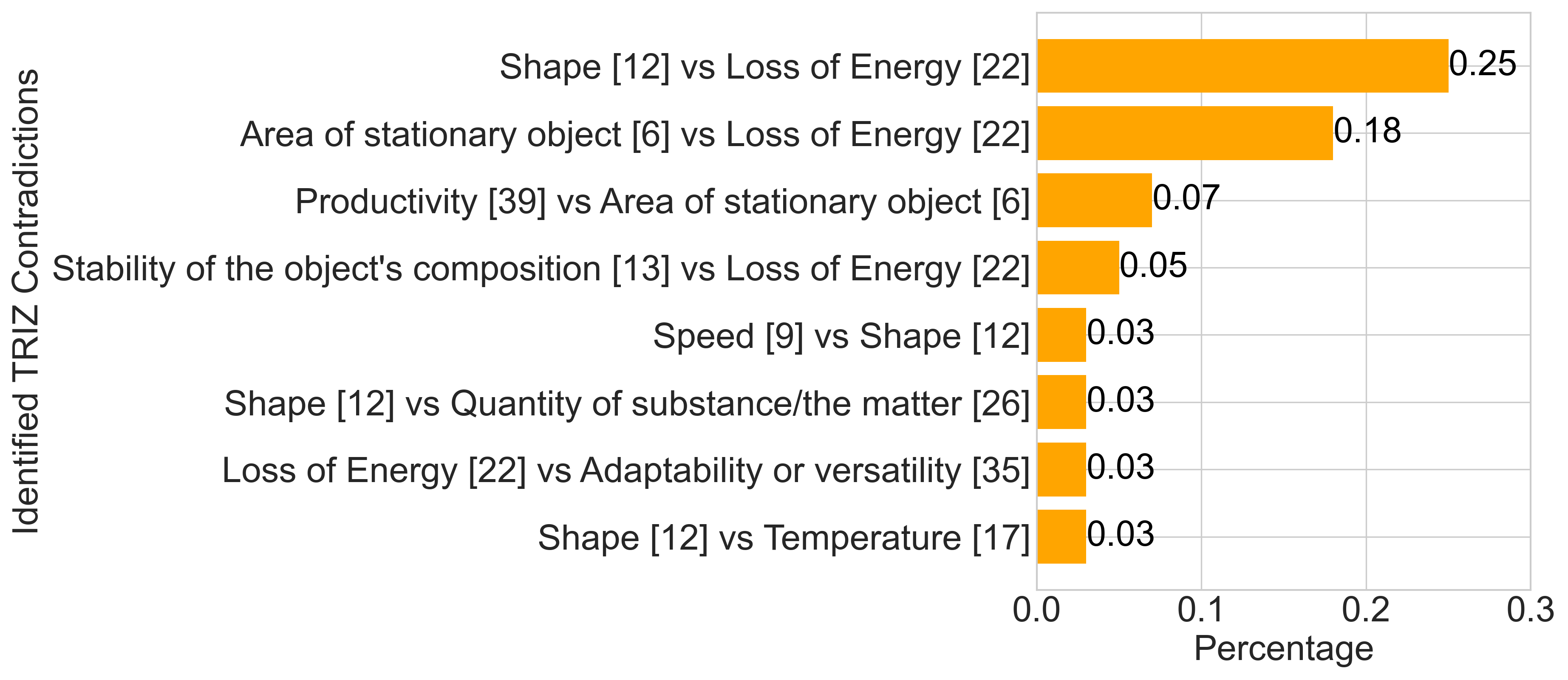}
	\caption{The frequency distribution of the identified engineering contradictions}
	\label{fig:fig6}
\end{figure}

The contradiction between \texttt{\{Shape (CP12)\}} and \texttt{\{Loss of Energy (CP22)\}} appeared most frequently, followed by \texttt{\{Area of stationary object (CP6)\}} vs \texttt{\{Loss of Energy (CP22)\}}. These two pairs of contradictions appeared significantly more often than others in the total 100 trials, reflecting the main trade-offs of this problem, specifically between the cooling requirements of the batteries and the contact area determined by the shape of the batteries and the heat pipes.

We randomly selected one report from the two contradictions described above. The reports are shown in Figure 7\footnote{Report A corresponds to the engineering contradiction with the highest frequency in Figure 6; Report B corresponds to the engineering contradiction with the second-highest frequency in Figure 6.}.

\begin{figure}[H]
	\centering
	\includegraphics[width=16cm]{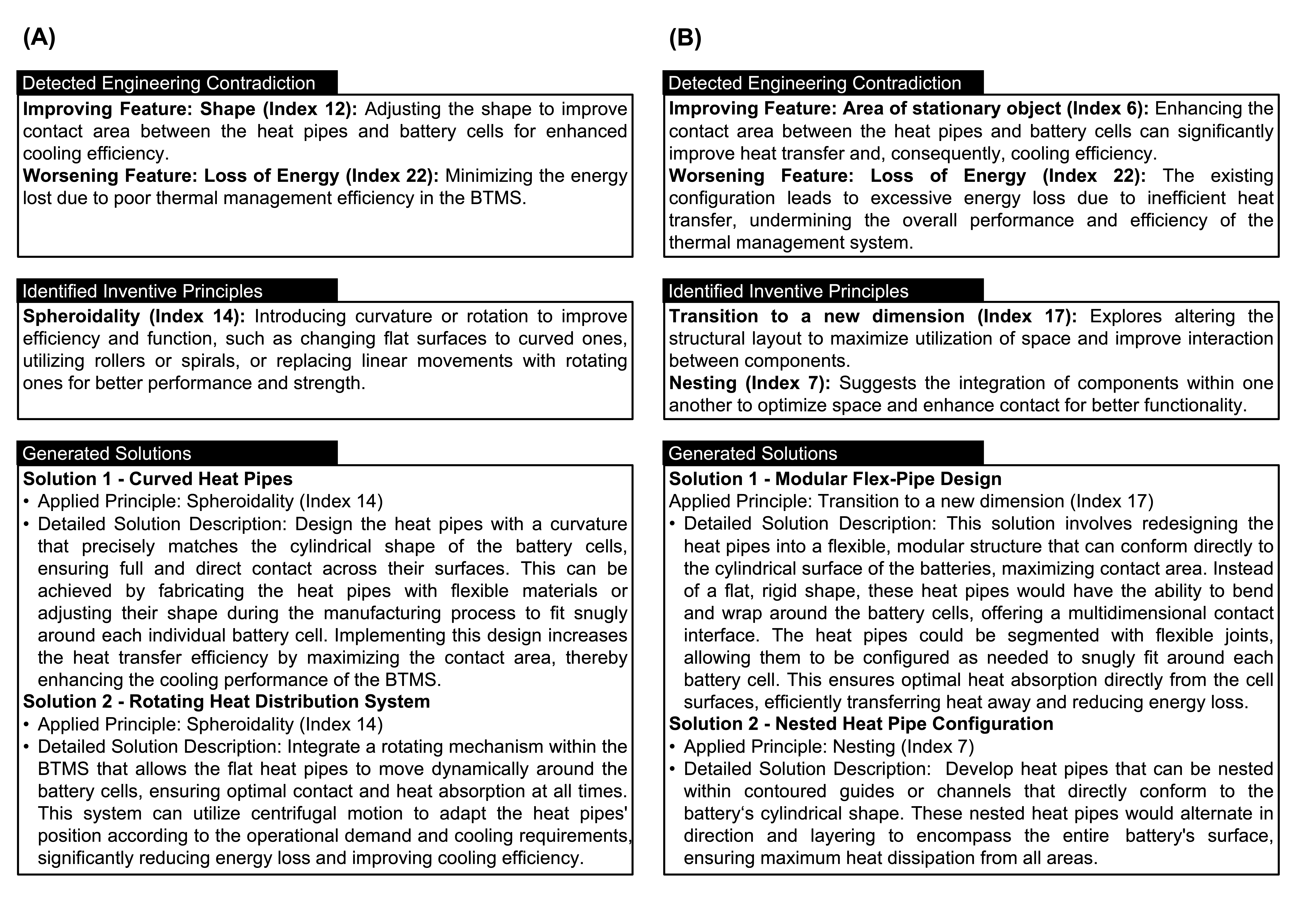}
	\caption{Two AutoTRIZ-generated solution reports for the BTMS Design Problem}
	\label{fig:fig7}
\end{figure}

Report A identified the inventive principle of \texttt{\{Spheroidality (IP14)\}} through the contradiction between CP12 and CP22. In this report, AutoTRIZ proposed using curved heat pipes (Solution 1) or rotating elements for optimal heat pipe-battery contact (Solution 2). These AutoTRIZ-generated solutions are similar to the approaches found in existing effective schemes, such as using Aluminum Sleeves \cite{Wang2019} and Conduction Elements \cite{Gan2020}. Figures 8A and 8B illustrate two sintered heat pipe-based BTMS schemes. In these designs, the evaporation section of the heat pipe is embedded in the conductive structures, indirectly contacting the battery.

In contrast, the condenser section extends into a liquid cooling tank or air-cooling channel. These methods have been proven effective in controlling battery temperature under different discharge rates \cite{Wang2019,Gan2020}. The AutoTRIZ-generated solutions align well with the field's current development trends. However, there is still room for further optimization to achieve high energy density and lightweight design.

Report B identified the contradiction between CP6 and CP22, suggesting the inventive principles of \texttt{\{Nesting (IP7)\}} and \texttt{\{Transition to a New Dimension (IP17)\}}. By applying these principles, AutoTRIZ generated the following specific solutions: (1) Develop heat pipes that can be nested within contoured guides or channels that directly conform to the battery’s cylindrical shape. (2) Redesign the heat pipes into a flexible, modular structure that can conform directly to the cylindrical surface of the batteries, maximizing contact area.

These two solutions focus on altering the form of the heat pipes. We can refer to the first solution to design heat pipes embedded in guides or channels shaped according to specific forms (such as bionic nest structures or curved surfaces) and use modular designs to enhance the contact area between the heat pipes and the batteries.

Inspired by the above results, we integrated the heat pipes with auxiliary conductive structures and designed a novel flat heat pipe-based design BTMS (FHP-BTMS, Figure 8C). Unlike traditional sintered heat pipes, the proposed flat heat pipe has interconnected chambers, expanding the heat pipe-based battery thermal management from one dimension to two dimensions (refer to \texttt{\{Transition to a New Dimension (IP17)\}}). It also includes grooved structures designed for the battery support frame and contact surfaces designed explicitly for cylindrical batteries (refer to \texttt{\{Nesting (IP7)\}}). Such a structure allows efficient battery assembly without introducing redundant components. Heat dissipation fins are also implemented on the end of the entire structure, enhancing heat convection at the cooling section. As shown in Figure 8C, the overall design significantly improves system integration and structural reliability while also achieving the goals of high energy density.

\begin{figure}[H]
	\centering
	\includegraphics[width=16cm]{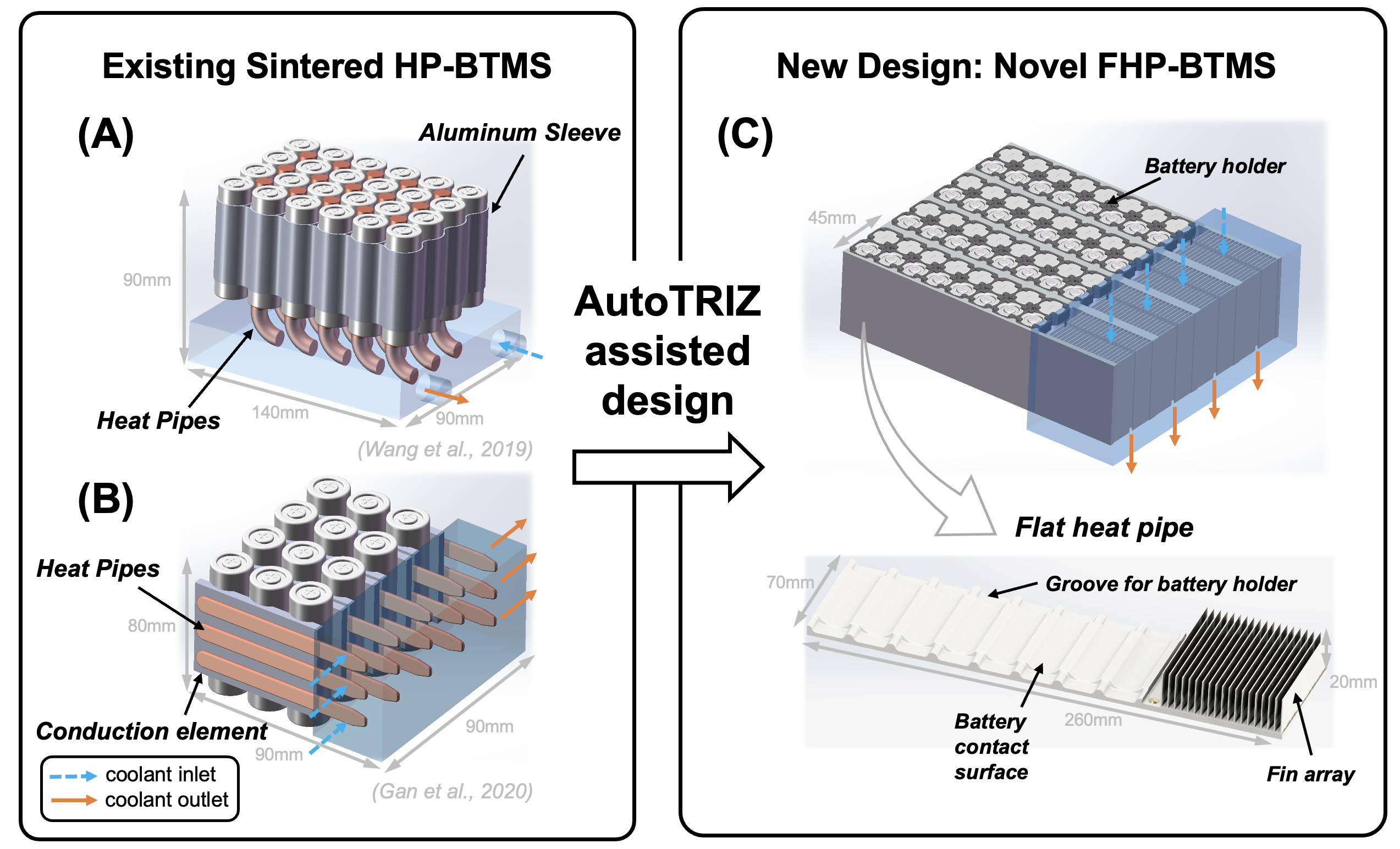}
	\caption{Existing sintered HP-BTMS (Schemes A and B) and the proposed novel FHP-BTMS with the assistance of AutoTRIZ (Scheme C)}
	\label{fig:fig8}
\end{figure}

\subsection{Design Evaluation through Computational Modeling}

In this Section, the proposed new FHP-BTMS design was evaluated. We used the following three metrics for evaluation: (1) grouping efficiency, (2) volumetric energy density, and (3) thermal management efficiency.

Grouping efficiency ($e_g$) is defined as the ratio of the battery volume ($V_{\text{batt}}$) to the total volume of the battery module ($V_{\text{module}}$), including thermal management components. It is calculated as follows:
$$e_g=V_{\text{batt}}/V_{\text{module}}$$

Volumetric energy density ($\text{SE}_V$) is calculated as the energy contained per unit volume of the battery module:
$$\text{SE}_V=E_{\text{batt}}/V_{\text{module}}$$

Thermal management efficiency ($e_{\text{{TMS}}}$) is the amount of battery heat generation dissipated per unit of thermal management energy consumption. It is calculated as follows:
$$e_{\text{TMS}} = \frac{\int Q\,dt - cm_{\text{batt}}(T_{\text{end}} - T_0)}{\int P\,dt}$$

where $c$ is the specific heat capacity of the battery, $m_{\text{batt}}$ is the mass of the battery, $T_{\text{END}}$ and $T_0$ represent the initial and final temperatures of the battery during discharge respectively. The battery heat generation rate $Q$ can be calculated using the simplified formula proposed by Xie et al. \cite{Xie2023}: $Q = I^2 R(\text{soc})$, where $I$ is the current flowing through each cell, and $R(\text{soc})$ is the internal resistance of the battery, which varies with the state of charge (soc). The power of the thermal management system $P$ can be calculated as: $P = \frac{\rho q^{3/2}}{S^2 \eta}$, where $\rho$ is the density of the cooling medium, $S$ is the cross-sectional area of the flow channel, $q$ is the coolant flow rate, and $\eta$ represents the efficiency of the thermal management components. We set $\eta = 80\%$ in this study. Appendix 2 describes more details of the calculations.

Table 1 lists all the parameters required for computing three evaluation metrics and the evaluation results of each BTMS scheme. In this table, Scheme (A) and Scheme (B) represent existing sintered HP-BTMS (also shown in Figure 8A and 8B \cite{Wang2019,Gan2020}), while Scheme FHP-BTMS represents our proposed novel design in this study (also shown in Figure 8C). Compared with existing schemes, the proposed novel design can improve grouping efficiency and volumetric energy density by 20-30\% while increasing thermal management efficiency by more than 60\%. We have shown the improvements compared to the best-performing existing scheme for each metric in Table 1. Figure 9A visualizes the comparison results of different BTMS schemes using a radar chart with three metrics. The radar chart is normalized with the maximum value for each dimension set to 100\%. Since FHP-BTMS achieves the best performance in all three dimensions, this figure effectively shows the relative performance of Scheme (A) and Scheme (B) compared to FHP-BTMS in each dimension.

\begin{table}[htbp]
\centering
\caption*{Table 1: Evaluation Results} % 手动编号 Table 1
\renewcommand{\arraystretch}{1.3}
\setlength{\tabcolsep}{3pt} % 控制列间距，避免超页边
\begin{tabular}{c ccccccc ccc}
\toprule
\textbf{Schemes} & \multicolumn{6}{c}{\textbf{Modeling Parameters}} & \multicolumn{3}{c}{\textbf{Metrics}} \\
\cmidrule(lr){2-7} \cmidrule(lr){8-10}
& $V_{\text{batt}}$ & $V_{\text{module}}$ & $T_{\text{END}} - T_0$ & $q$ & Coolant & $S$ & $SE_V$ & $e_{\text{TMS}}$ & $e_{g}$ \\
& (L) & (L) & ($^\circ$C) & (L/min) & & (cm$^2$) & (Wh/L) & & \\
\midrule
\textbf{Scheme (A) \cite{Wang2019}} & 0.40 & 1.16 & 7 & 2 & Water & 0.78 & 166 & 13.18 & 34\% \\
\textbf{Scheme (B) \cite{Gan2020}} & 0.20 & 0.61 & 2.5 & 2 & Water & 21.60 & 158 & 11.03 & 33\% \\
\textbf{FHP-BTMS (Ours)} & 0.39 & 0.91 & 7.5 & 1950 & Air & 32.50 & 206 & 21.66 & 43\% \\
\textbf{Improvements} & & & & & & & \textbf{+24\% ↑} & \textbf{+64\% ↑} & \textbf{+26\% ↑} \\
\bottomrule
\end{tabular}
\end{table}

Figure 9B compares our proposed scheme (air cooling with FHP, referred to as FHP below) with two thermal management modes (i.e., air cooling and liquid cooling) under the same discharge and cooling power conditions. The results show the Max Temperature Rise (°C) and Max Temperature Difference (°C) for the three cooling modes. The cooling fluid velocities for FHP, air cooling, and liquid cooling are 10 m/s (air), 2 m/s (air), and 0.005 m/s (water), respectively.

The battery thermal management performance with air cooling alone is significantly lower than that of FHP and liquid cooling. FHP and liquid cooling exhibit similar thermal management effects in controlling the temperature rise. Besides, the temperature differences of the batteries based on FHP is the best among all modes. With a nearly uniform heat source power input on the evaporator side of the heat pipes, FHP can achieve uniform heat dissipation for each heat source, significantly reducing the temperature gradient between the cells.

\begin{figure}[H]
	\centering
	\includegraphics[width=12cm]{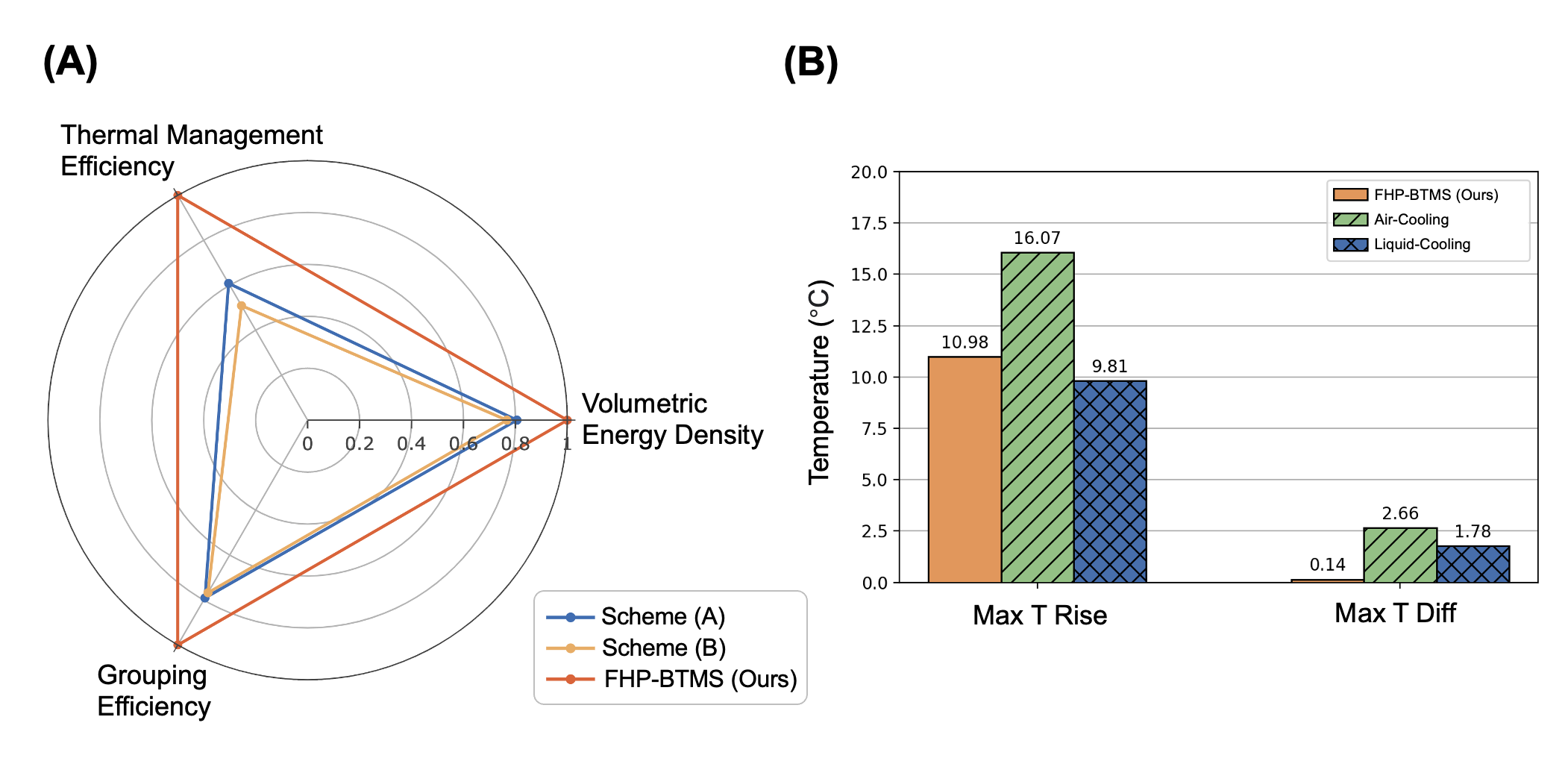}
	\caption{(A) Comparison of different schemes on three performance metrics; (B) Comparison of different cooling modes.}
	\label{fig:fig9}
\end{figure}

\subsection{Design Optimization through Parameter Changes}

In the solution reports, we also noted that through the third most frequently identified engineering contradiction (refer to Figure 6) between \texttt{\{Productivity (CP39)\}} and \texttt{\{Area of stationary object (CP6)\}}, AutoTRIZ finds the inventive principle of \texttt{\{Transforming the physical or chemical state of an object (IP35)\}}. Our BTMS scheme could employ parameter changes to achieve a more optimized design. For example, applying phase-change materials for the heat pipe that can adapt its shape to achieve complete contact; or optimizing the physical parameters related to the enveloping surface.

In the proposed design, the heat pipes with cylindrical envelopment surfaces closely adhere to the batteries, ensuring efficient heat dissipation. Inspired by the suggestions generated by AutoTRIZ, we consider further adjusting the geometric design parameters of the envelopment surfaces. As illustrated in Figure 10A, we focus on the contact angle $\theta$ (between the flat heat pipe and the cylindrical battery) as a critical parameter for optimization. We then compared the thermal management effects at different contact angles.

Figure 10B shows how the final maximum temperature of the battery module changes under different discharge rates as the $\theta$ varies\footnote{The heat generation rates of the battery under different discharge rates are listed in Appendix 2.}. All conditions start from an initial battery temperature of 20°C. We can see that under the extreme discharge condition of 3C, a contact angle of at least 20° is required to keep the battery temperature below 45°C, widely considered the upper limit for the regular operation of lithium-ion batteries. Increasing $\theta$ can cause a rapid increase in the wall thickness of the flat heat pipe, leading to unnecessary manufacturing costs and difficulties. Therefore, in our final design, selecting a contact angle $\theta$ from 20° to 45° can achieve optimal performance under various discharge conditions.

\begin{figure}[H]
	\centering
	\includegraphics[width=14cm]{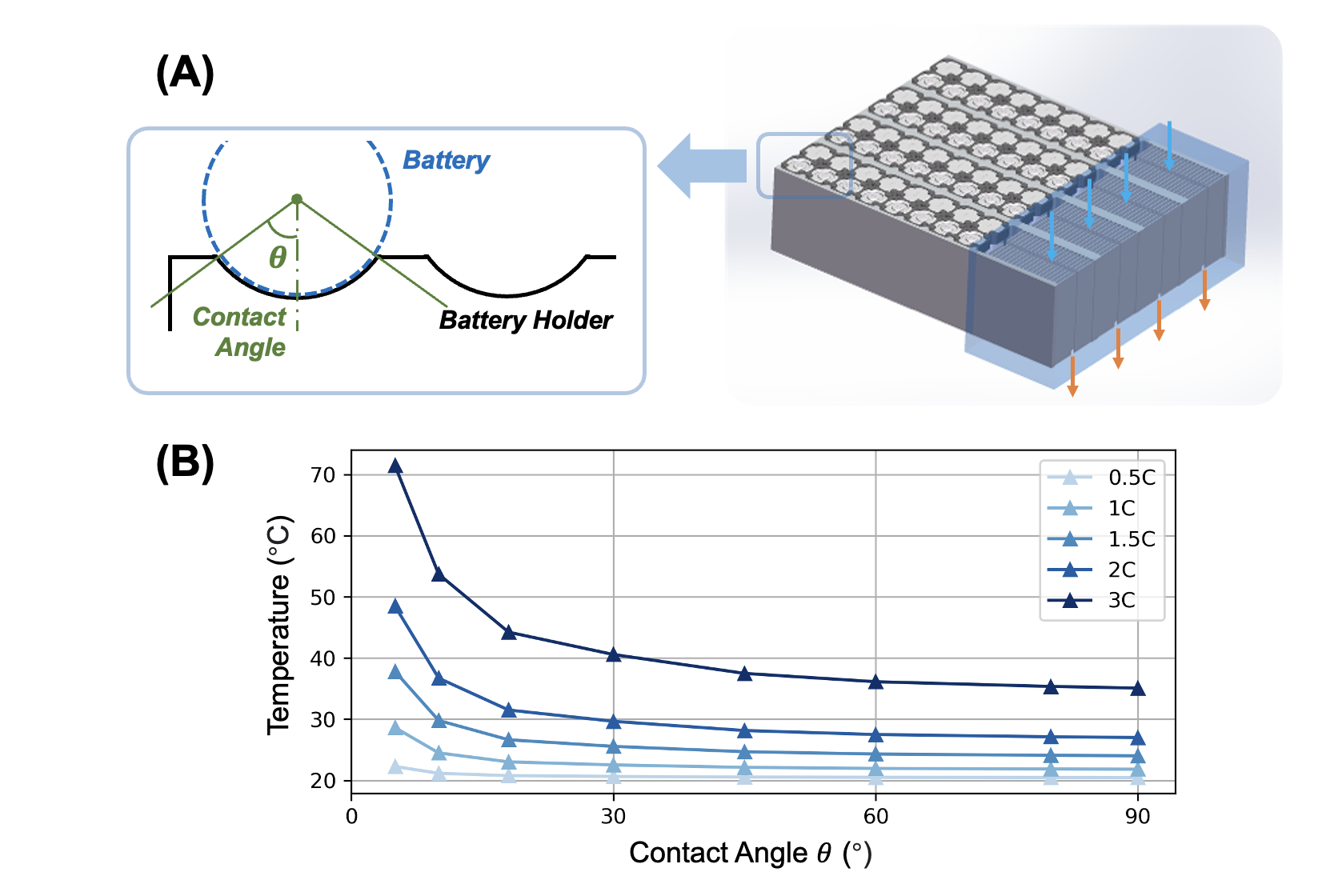}
	\caption{(A) The illustration of Contact Angle; (B) Temperature Variation with Different Contact Angle under Different Discharge Rates}
	\label{fig:fig10}
\end{figure}

In brief, the above case study of BTMS design demonstrates that AutoTRIZ can generate novel and valuable design solutions in the engineering field. Throughout the entire design process, AutoTRIZ can help engineers follow TRIZ reasoning logic to identify innovative problem-solving directions and provide insightful design information embedded in generated AutoTRIZ reports. Such human-AI collaboration can significantly enhance design creativity, improve design efficiency, and ensure more robust and optimized design outcomes.

\section{Discussion}
\label{sec:sec6}

So far, we have presented AutoTRIZ, a novel artificial ideation tool that integrates LLMs and TRIZ to automatically generate inventive solutions for any given problem in an interpretable way. We have demonstrated AutoTRIZ’s effectiveness and practicality through comparative experiments with textbook cases and the BTMS case study. To use AutoTRIZ, users only need to provide a problem statement in descriptive text, and AutoTRIZ will follow the TRIZ reasoning flow to generate a solution report automatically. In practice, one can extract insightful information from the generated solutions or use them directly for further problem-solving.

In the BTMS design case, AutoTRIZ automatically identified several inventive principles of TRIZ and applied them to generate solutions for effective temperature control and lightweight structure. Specifically, one AutoTRIZ report used \texttt{\{Spheroidality (IP14)\}} to suggest the use of curved heat pipes or rotating elements as a solution, which aligns with existing sintered heat pipe-based BTMS schemes in the field \cite{Wang2019,Gan2020}. This demonstrates AutoTRIZ's ability to incorporate advanced engineering knowledge and problem-solving skills in this domain. Another AutoTRIZ report employed the principles of \texttt{\{Nesting (IP7)\}} and \texttt{\{Transition to a New Dimension (IP17)\}} to recommend considering nested pipes and flat heat pipes with interconnected chambers in the design, guiding us to develop a novel scheme (FHP-BTMS). Further utilizing the inventive principle of \texttt{\{Transforming the physical or chemical state of an object (IP35)\}}, we optimized the contact angle between the flat heat pipe and the battery to achieve better thermal management effects. During the use of AutoTRIZ, the input information only includes the problem background, the issues with the current scheme, and the desired design goals. Computational modeling results show that the design assisted by AutoTRIZ outperforms grouping efficiency, volumetric energy density, and thermal management efficiency compared to existing schemes.

In this case, AutoTRIZ found the primary engineering contradiction between shape/contact area and energy loss, which aligns with the experts' understanding of problem\footnote{One of the co-authors of this paper is an expert in this field and is responsible for verifying results generated by AutoTRIZ.}. Compared to the manual use of traditional TRIZ, engineers in the battery thermal management field no longer need to master all definitions of TRIZ contradiction parameters, the contents of inventive principles, and the skills to apply these principles to solve such specific problems. The generated reports include detailed reasoning processes and explanations for each step, enabling engineers to read the report, conduct minimal necessary checks on the professional information, and either draw inspiration or directly apply the solutions generated to solve the problem. In contrast, directly using vanilla LLMs (such as ChatGPT) for problem-solving requires engineers to provide more detailed information while describing the engineering problem. In the case study, we also tried to guide ChatGPT to use TRIZ for solution generation directly using its chat interface. However, the TRIZ reasoning results always included apparent errors, such as mismatched contradictions and inventive principles. Therefore, AutoTRIZ fills this gap by providing a more reliable, accurate, and interpretable solution generation process that seamlessly integrates TRIZ principles and LLMs’ capabilities, ensuring the generated solutions are innovative and practically applicable to real-world engineering problems.

Generally, AutoTRIZ can solve engineering problems across multiple domains. Our case base of 10 problems in this study spans diverse fields, and AutoTRIZ has effectively generated inventive solutions for each case. We further explored its capabilities through case studies, showcasing its effectiveness in these areas. In the engineering field, prior studies \cite{Picard2023,Makatura2023} have assessed the capabilities of LLMs across a broad range of engineering-related tasks, revealing that these models have extensive engineering knowledge, such as in design and manufacturing. Therefore, in our framework, we only control the reasoning flow without limiting the knowledge involved in the ideation process to fully leverage LLMs' general expertise and capabilities. Compared with other LLM-based ideation tools, such as InnoBID \cite{Zhu2023bio} and AskNatureNet \cite{Chen2024AskNatureNet}, AutoTRIZ has a key advantage in its integration of the TRIZ methodology. By generating a complete TRIZ reasoning process, AutoTRIZ enables users to understand the specific-to-general-to-specific problem-solving approach. This structured reasoning not only improves interpretability and transparency but also helps users trace the logical steps behind the generated solutions. However, AutoTRIZ is also constrained by the limitations of TRIZ databases. For instance, the original TRIZ defines only 40 inventive principles, which may not fully capture the vast diversity of modern engineering and innovation challenges. Combining AutoTRIZ with other LLM-based ideation methods could help address this limitation by expanding the range of possible solutions, incorporating broader design knowledge, and adapting TRIZ-based reasoning to a wider variety of problem domains.

Besides, in the comparative study presented in Section 4.3, we observed that the problem statement contains information related to the desired direction of improvement, which is relevant to the contradiction. Such information aids in aligning AutoTRIZ’s detections with those of human experts. Accordingly, as demonstrated in Figure 11, we can incorporate multi-input configurations into the system, enabling AutoTRIZ to generate solutions that fully consider detailed requirements from users. Additionally, integrating structured user inputs, such as specifying key constraints or preferred solution attributes, could further enhance the relevance and practicality of the generated solutions. We keep it simple to ensure accessibility for all users, including those without an understanding of TRIZ. We plan to investigate user interaction with TRIZ, AutoTRIZ, and vanilla LLMs, examining the differences to identify the most effective methods for improving the overall user experience and system performance.

\begin{figure}[H]
	\centering
	\includegraphics[width=16cm]{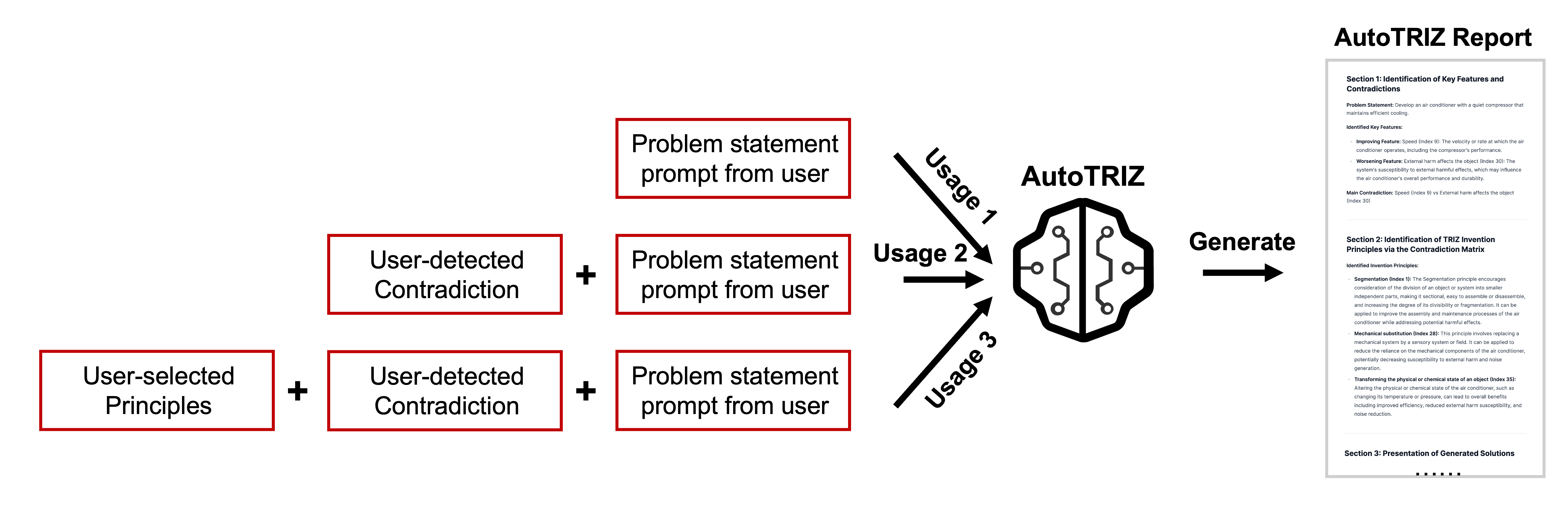}
	\caption{The multi-input usages of AutoTRIZ}
	\label{fig:fig11}
\end{figure}

Overall, this work has made at least three contributions to the field of data-driven engineering innovation \cite{Vlah2022,Luo2022,Jiang2021}. First, the proposed AutoTRIZ significantly reduces the entry barrier to TRIZ. It can generate a multitude of solutions in a short period because it leverages the computational power and vast knowledge base of LLMs. Its user-friendly interface further enhances this efficiency, allowing for easy configuration and use, significantly reducing the time needed to generate ideas and refine problem-solving strategies. In contrast, mastering the traditional TRIZ method for professional use typically requires months of training and substantial intellectual and cognitive efforts \cite{Ilevbare2013}.

The second contribution is using generative AI to generate innovative solutions in a controllable and interpretable way. Recent studies \cite{Zhu2023bio,Zhu2023gen} have demonstrated that LLMs can be used effectively for concept generation and diverse ideation. LLM-based tools possess extensive cross-domain knowledge, allowing them to break beyond pre-defined boundaries and explore a broader solution space \cite{Chang2023}. Similarly, TRIZ is inherently cross-disciplinary. By integrating TRIZ with the reasoning capabilities of LLMs, AutoTRIZ provides a structured approach to ideation, ensuring that the reasoning process follows a clear and systematic path rather than being entirely unpredictable. The generated solutions are derived through a systematic framework, making the process more transparent and allowing engineers to understand not only what solutions are proposed but also why they are appropriate. In practice, users can directly implement the generated solutions or consider the provided information to redesign based on their preferences, fostering human-machine collaborative innovation \cite{Song2024}.

Thirdly, although this study focuses on automating the TRIZ reasoning process using LLMs, the proposed framework can be extended to automate other knowledge-based innovation methods. For instance, Yilmaz et al. \cite{Yilmaz2016} identified 77 design heuristics from over 3,000 design process outcomes and suggested a subset of these heuristics to designers. When these heuristics were randomly selected, they produced improved design outcomes \cite{Daly2012}. Since our framework is designed to integrate structured knowledge bases with LLM-driven reasoning, it has the potential to systematically organize, retrieve, and apply these heuristics in a controlled manner. By applying our framework to this research, one could treat the identified heuristics as an internal knowledge base for the LLM-based agent, determining how to utilize these heuristics in the backend. Moreover, to develop a more powerful tool, one could also integrate various knowledge-based idea generation methods into the reasoning modules of LLMs, such as SCAMPER \cite{Eberle1996}, IDEO Method Cards \cite{ideo2003}, Bio-inspired Design \cite{fu2014}, and Design-by-Analogy \cite{Jiang2021,murphy2014function,hey2008analogies}.

\section{Limitation and Future Work}

The proposed AutoTRIZ framework has several limitations. Firstly, the solutions generated by LLMs may contain hallucinations or erroneous information \cite{Jiang2024}. Future work may include fact-check modules to ensure the accuracy of the solutions. Additionally, there is no objective mechanism to evaluate the effectiveness of generated solutions. Users must independently assess solution quality and rank them for practical use. In the solution generation step, we can also incorporate the latest advancements in LLM applications to guide reasoning. For example, multiple solutions could be generated for re-ranking to provide higher-quality results to users \cite{Jiang2023Blender}. Besides, iterative results optimization could be implemented using LLMs as optimizers before generating the final report \cite{Yang2024Optimizers}. The current tool identifies and resolves a single contradiction at a time, which may limit the exploration of alternative contradictions that could lead to different innovation pathways. Future improvements could focus on enabling multi-perspective contradiction analysis.

The current paper primarily demonstrates the procedures and effectiveness of using AutoTRIZ to aid engineering problem-solving. We have done so through comparative experiments with textbook cases and a detailed BTMS design case study. In further work, we plan to involve more experts with diverse backgrounds in analyzing the same problems in terms of novelty, usefulness, and breadth of the generated solutions for comparison, making the conclusions more robust. Furthermore, this study was demonstrated on a limited set of problem cases, providing only an initial insight into AutoTRIZ that might introduce some bias. In future research, we aim to apply this method to a broader and more diverse range of problems, systematically evaluating AutoTRIZ's performance. In addition, exploring comparisons between AutoTRIZ and other AI-driven TRIZ methods is also a valuable direction for future research.

Moreover, a more fundamental understanding of the science behind leveraging LLMs for innovation is needed, including the boundaries of LLMs' capabilities in problem-solving and concept generation and the dynamics of content generation for these engineering-related tasks. These insights can help us better apply LLMs within the proposed framework. Also, introducing multimodal capabilities into the current model may be worth exploring \cite{Fei2022}.

Additionally, we recognize that AutoTRIZ currently requires users to have some level of domain-specific knowledge to interpret the generated solutions effectively. Future work will focus on improving accessibility by incorporating more intuitive explanations and interactive guidance, assisting non-experts better understand and utilize the tool.

\section{Conclusion}

This paper proposes AutoTRIZ, an artificial ideation framework and system that leverages LLMs to automate the TRIZ methodology and enhance its applications. AutoTRIZ is constructed by multiple LLM-based reasoning agents that interact with the inner fixed knowledge base. It takes problem statements from users as initial inputs and automatically generates an interpretable solution report by following the step-by-step TRIZ reasoning process. The effectiveness of AutoTRIZ is demonstrated via a series of comparative experiments with textbook cases and a detailed case study about battery thermal management system design. Although this paper primarily focuses on integrating LLMs with TRIZ, the proposed framework holds the potential to be extended to other knowledge-based ideation methods, including SCAMPER, Design Heuristics, and Design-by-Analogy.

\section*{Disclosure statement}

No potential conflict of interest was reported by the author(s).

\bibliographystyle{unsrtnat}
\bibliography{references}

\newpage
\section*{Appendix 1: Full Prompts and Function Implementation}

In Sections A1.1 and A1.2, we present the full prompts used in each LLM-driven module within AutoTRIZ. In Section A1.3, we provide the key function implementation for LLM-based function calls, enabling the retrieval of TRIZ inventive principles.

\subsection*{A1.1: System prompts}
\begin{tcolorbox}[breakable]
You are the world's leading engineering innovator and TRIZ expert, highly skilled in applying TRIZ theory and methods to solve challenging problems. When given a problem statement, you must strictly follow the processes outlined below to solve it. Remember to call getTrizPrinciple() in Module 3.
\end{tcolorbox}

\subsection*{A1.2: Full prompts}
\textbf{Module 1:}
\begin{tcolorbox}[breakable]
Your task is to identify the core problem from the input and express it as a clear and concise problem statement while filtering out unnecessary details such as background information, scenario descriptions, and redundant content.\\
Guidelines:\\
- Focus on extracting the key issue that needs to be solved.\\
- If the input is already a well-structured problem statement, keep it as is.\\
- If the input contains extra information, refine and summarize it into a direct problem statement.\\
- Ensure clarity and conciseness, making it easy for further TRIZ-based processing.
\end{tcolorbox}

\textbf{Module 2:}
\begin{tcolorbox}[breakable]
Transform the problem statement into a contradiction by selecting the improving parameter and the worsening parameter relevant to the problem. Carefully review the content step by step to ensure a clear understanding.\\
Here is the parameter list: \{Engineering\_Parameters\}.\\
Here are some examples: \{Examples\}.
\end{tcolorbox}

\textbf{Module 3:}
\begin{tcolorbox}[breakable]
Identify the relevant inventive principles using getTrizPrinciple().
\end{tcolorbox}

\textbf{Module 4:}
\begin{tcolorbox}[breakable]
Apply the relevant principles obtained from getTrizPrinciple() to solve the given problem. When generating solutions, ensure they are exceptionally comprehensive, covering the following aspects:\\
- Each solution should be articulated in great detail, explaining how it directly addresses the specific aspects of the problem. This should include a clear explanation of the mechanisms or processes involved in each solution.\\
- For each solution, provide a detailed step-by-step plan that outlines how it can be practically implemented. This should include necessary resources, estimated timelines, and any required steps or stages in the implementation process.\\
- Clearly demonstrate how each solution aligns with the TRIZ principles identified. Explain the rationale behind choosing each principle for the specific aspect of the problem it addresses.\\
- Identify any potential challenges or obstacles that might arise during the implementation of each solution and propose strategies to overcome them.\\
- Suggest methods for evaluating the effectiveness of each solution. Include specific metrics or criteria that can be used to assess whether the solution is working as intended.\\
- Provide an analysis of the expected benefits versus the costs (both financial and otherwise) associated with each solution. This should help in understanding the feasibility and value of the solutions proposed.
\end{tcolorbox}

\textbf{Module 5:}
\begin{tcolorbox}[breakable]
Summarize all the given content and format the output in Markdown according to the following structure: \{Output\_Format\_Example\}.
\end{tcolorbox}

\subsection*{A1.3: Function Implementation}

The following function (in TypeScript code) is implemented for LLM function calls, enabling the retrieval of TRIZ inventive principles based on the specified improving and worsening indices from the Contradiction Matrix.

\begin{tcolorbox}[breakable]
\begin{lstlisting}
async function getTrizPrinciple(improving_index: number, worsening_index: number) {
  const trizMatrix = [[null, [15, 8, 29]], [null, null, [10, 29, 15]]];

  const trizPrinciples = {
    "1": {"Title": "Segmentation", "Explanation": "..."},
    "2": {"Title": "Extraction", "Explanation": "..."},
    "3": {"Title": "Local Quality", "Explanation": "..."}
  };

  improving_index -= 1;
  worsening_index -= 1;

  if (improving_index >= trizMatrix.length || worsening_index >= trizMatrix.length) {
    return "Index out of range";
  }

  const principleIndices = trizMatrix[improving_index][worsening_index];

  if (!principleIndices) {
    return "No principle found for this case";
  }

  return principleIndices.map(index => trizPrinciples[index.toString()] || "Unknown principle");
}
\end{lstlisting}
\end{tcolorbox}

\newpage
\section*{Appendix 2: Calculation Details for BTMS Case Study}

In the case study presented in Section 5, we adopt a thermal network computational model to consider the coupled effects of multi-phase heat transfer within the FHP and heat generation from the battery. The temperature variation at each time step is determined by solving the energy conservation equations for the nodes shown in Figure A1, along with the thermal resistances connecting these temperature nodes. All the formulas for thermal resistance calculations are provided in Table A1, and the detailed explanation can be found in our previous work \cite{Dan2022}. The thermophysical properties and structural parameters required for the calculations are presented in Table A2 for the battery and Table A3 for the flat heat pipe.

The energy conservation equations governing each node in the thermal network are expressed as follow:

$$
c_i \rho_i V_i \frac{dT_i}{dt} = \left[ R_E^{-1} \quad R_W^{-1} \quad R_S^{-1} \quad R_N^{-1} \right] 
\left( 
\begin{bmatrix}
T_E \\
T_W \\
T_S \\
T_N
\end{bmatrix} 
- T_i 
\right) + Q_i
$$

where E (east), W (west), S (south), N (north) are the directions of the control volumes surrounding volume $i$.

In addition, to evaluate the proposed designs in Section 5.4, we used different discharge rates to analyze their impact on thermal performance. The heat generation rates of the battery under different C-rates are listed in Table A4, which were used as the input to calculate the corresponding temperature variations (Figure 10B).

\begin{figure}[H]
	\centering
	\includegraphics[width=10cm]{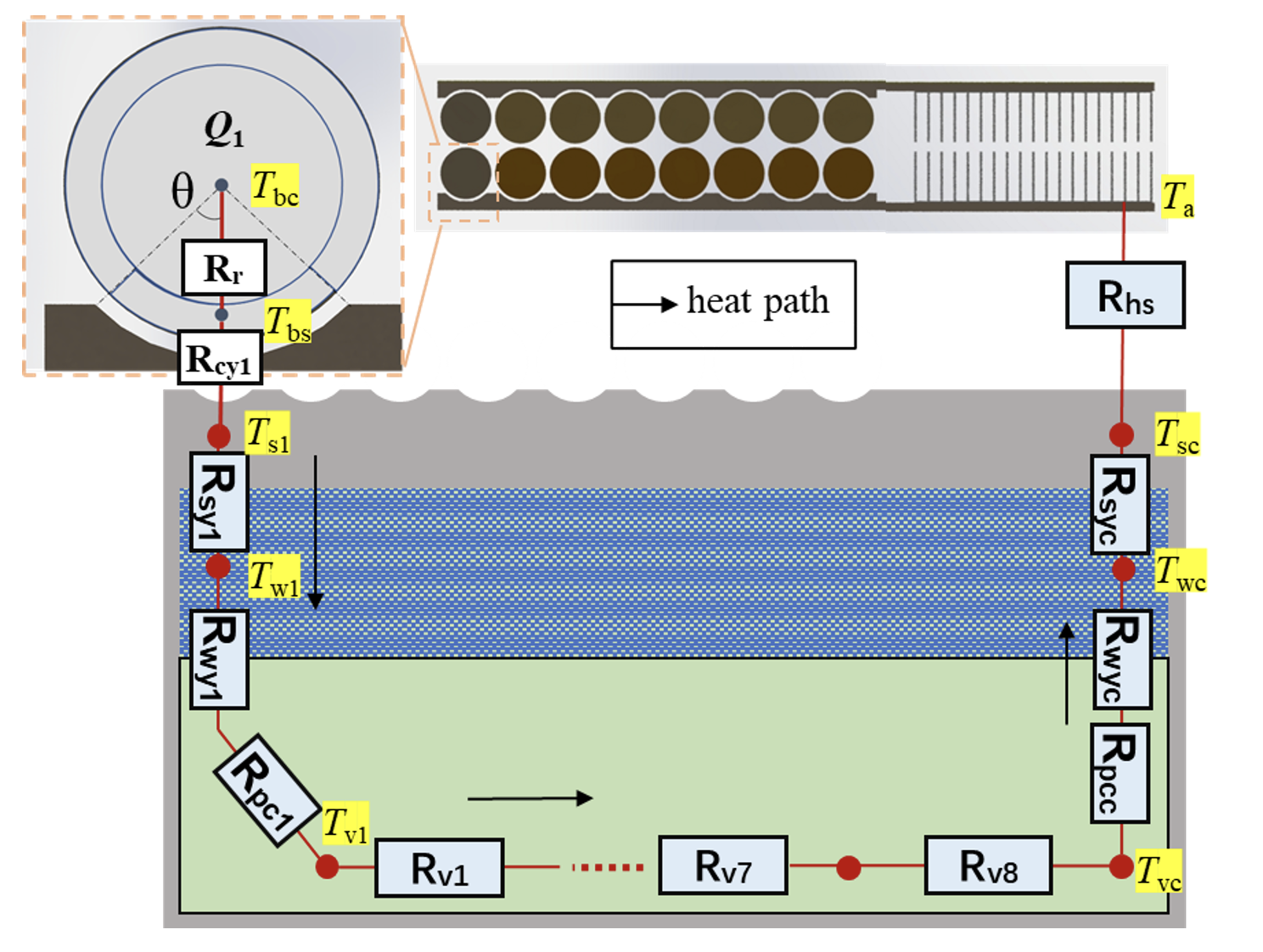}
	\caption*{Figure A1: Thermal network computational model of FHP-BTMS}
	\label{fig:figA1}
\end{figure}

\begin{table}[htbp]
\centering
\caption*{Table A1: Thermal resistance models}
\renewcommand{\arraystretch}{1.2}
\begin{tabularx}{\textwidth}{>{\centering\arraybackslash}X >{\centering\arraybackslash}X}
\toprule
\textbf{Type} & \textbf{Expression} \\
\midrule
Radial thermal resistance & $R_r = \int \frac{dr}{k_r \cdot 2\theta r_b}$ \\
Connection thermal resistance & $\frac{1}{R_{ci}} = \int_{-\theta}^{\theta} \frac{k_s r_b h_b}{t_s + r_b (1 - \cos \varphi)} \, d\varphi$ \\
Phase change thermal resistance\cite{Dan2022} & $R_{pci} = \frac{2 - \sigma \left( 2 \pi R_{\text{gas}} T_{vi} \right)^{0.5} R_{\text{gas}} T_{vi}^2}{2 \sigma A_i p_{vi} H_{fgi}^2}$ \\
Vapor flow thermal resistance\cite{Dan2022} & $R_{vi} = \frac{R_{\text{gas}} T_{vi}^2}{p_{vi} h_{fgi}} \cdot \frac{12 \mu_{vi}}{t_v^2 \rho_{vi} A_i H_{fgi}} \, l_i$ \\
Heat sink thermal resistance & $R_{hs} = \frac{1}{h_{\text{fin}} A_{\text{fin}}}$ \\

\bottomrule
\end{tabularx}
\end{table}

\begin{table}[htbp]
\centering
\caption*{Table A2: Basic information of the 21700 cylindrical battery}
\renewcommand{\arraystretch}{1.2}
\begin{tabularx}{\textwidth}{>{\centering\arraybackslash}X >{\centering\arraybackslash}X}
\toprule
\textbf{Parameter} & \textbf{Value} \\
\midrule
Capacity & 4.8 Ah \\
Mass & 0.069 kg \\
Material & cathode: graphite, anode: NCM \\
Nominal voltage & 3.65 V \\
Dimension & $\phi$21.7mm $\times$ 70mm \\
Specific heat capacity & 920 J/(kg·K) \\
Thermal conductivity (radial direction) & 1.38 W/(m·K) \\
Thermal conductivity (other direction) & 32.52 W/(m·K) \\
\bottomrule
\end{tabularx}
\end{table}

\begin{table}[htbp]
\centering
\caption*{Table A3: Physical properties and structural parameters of the flat heat pipe}
\renewcommand{\arraystretch}{1.2}
\begin{tabularx}{\textwidth}{
  >{\centering\arraybackslash}X
  >{\centering\arraybackslash}m{2.5cm}
  >{\centering\arraybackslash}X}
\toprule
\textbf{Parameter} & \textbf{Symbol} & \textbf{Value} \\
\midrule
Shell Material & -- & Aluminum \\
Wick Material & -- & (Porous sintered) Aluminum particles \\
Working Fluid & -- & Acetone \\
Evaporator length & $l_{ei}$ & 23mm per cell \\
Condenser length & $l_c$ & 65mm \\
FHP width & $w_{\text{FHP}}$ & 70mm \\
FHP length & $l_{\text{FHP}}$ & 260mm \\
Total thickness of FHP & $t_{\text{FHP}}$ & 4.2mm \\
Thickness of shell & $t_s$ & 1mm \\
Thickness of wick & $t_w$ & 1.1mm \\
Thickness of vapor channel & $t_v$ & 1.1mm \\
Thickness of fin & $t_{\text{fin}}$ & 0.6mm \\
Width of fin & $w_{\text{fin}}$ & 20mm \\
Spacing of fin & $s_{\text{fin}}$ & 3mm \\
Wick thermal conductivity & $k_w$ & 9.965 W/(m·K) \cite{Wang2023BatteryPipe} \\
Wick thermal capacity & $c_w$ & 1059 J/(kg·K) \cite{Wang2023BatteryPipe} \\
Shell thermal capacity & $c_s$ & 920.9 J/(kg·K) \\
Battery thermal capacity & $c_b$ & 1023 J/(kg·K) \\
Wick density & $\rho_w$ & 1059 kg/m$^3$ \cite{Wang2023BatteryPipe} \\
Battery density & $\rho_b$ & 2519 kg/m$^3$ \\
\bottomrule
\end{tabularx}
\end{table}

\begin{table}[htbp]
\centering
\caption*{Table A4: Heat generation rate of the battery under different C-rates}
\renewcommand{\arraystretch}{1.2}
\begin{tabularx}{\textwidth}{
  >{\centering\arraybackslash}X
  >{\centering\arraybackslash}m{1.5cm}
  >{\centering\arraybackslash}m{1.5cm}
  >{\centering\arraybackslash}m{1.5cm}
  >{\centering\arraybackslash}m{1.5cm}
  >{\centering\arraybackslash}m{1.5cm}}
\toprule
\textbf{C-rate} & 0.5 & 1 & 1.5 & 2 & 3 \\
\midrule
\textbf{Heat generation rate (W)} & 0.19 & 0.76 & 1.71 & 3.04 & 6.44 \\
\bottomrule
\end{tabularx}
\end{table}

\end{document}